%% file: ms.tex
\documentclass[10pt,journal,compsoc]{IEEEtran}
\usepackage{color, colortbl}
\usepackage[first=0,last=9]{lcg}

\usepackage{booktabs}
\usepackage{hyperref}
\usepackage[utf8]{inputenc}
\usepackage[T1]{fontenc}
\usepackage{paralist}
\usepackage{soul}
\usepackage{graphicx}
\usepackage{tabularx}
\usepackage{color}
\usepackage{footmisc}
\usepackage{balance}

\usepackage{subcaption}
\usepackage{pdflscape}
\usepackage{afterpage}
\usepackage[table]{xcolor}
\usepackage{multirow}
\usepackage{ltablex}
\usepackage{threeparttable}

\usepackage{adjustbox}
\usepackage{array}
\newcolumntype{Z}{>{\raggedright\arraybackslash}X}

\newcolumntype{P}[1]{>{\raggedright\arraybackslash}p{#1}}

\newcolumntype{R}[2]{%
  >{\adjustbox{angle=#1,lap=\width-(#2)}\bgroup}%
  l%
  <{\egroup}%
}
\newcommand*\rot{\multicolumn{1}{R{90}{1em}}}

\makeatletter

\def\ps@IEEEtitlepagestyle{%
  \def\@oddfoot{\mycopyrightnotice}%
  \def\@evenfoot{}%
}
\def\mycopyrightnotice{%
  {\scriptsize This work has been submitted to the IEEE for possible publication. Copyright may be transferred without notice, after which this version may no longer be accessible.}
  \gdef\mycopyrightnotice{}
}

\newcommand\footnoteref[1]{\protected@xdef\@thefnmark{\ref{#1}}\@footnotemark}
\makeatother

\usepackage[dot]{bibtopic}

\usepackage[colorinlistoftodos]{todonotes}

\definecolor{LightGray}{gray}{0.9}
\definecolor{Yellow}{HTML}{FFFFCC}

\ifCLASSOPTIONcompsoc
 \usepackage[nocompress]{cite}
\else
 \usepackage{cite}
\fi
\ifCLASSINFOpdf
\else
\fi

\usepackage{stfloats}
\usepackage{url}

\begin{document}
%
\title{Backsourcing of Software Development --- \\A Systematic Literature Review}
%
%
%
%

\author{Jefferson Seide Moll\'eri,
    Casper Lassenius,
    Magne Jørgensen
\IEEEcompsocitemizethanks{\IEEEcompsocthanksitem J. Moll\'eri
and M. Jørgensen are with Simula Metropolitan Centre for Digital 
Engineering, Norway.  C. Lassenius is with Aalto University, Finland 
and Simula Metropolitan Centre for Digital Engineering, Norway.
\protect\\
E-mail: jefferson@simula.no\

}
}

\IEEEtitleabstractindextext{%
\begin{abstract}
\textbf{Context:} Backsourcing is the process of insourcing previously outsourced activities. When companies experience environmental or strategic changes, or challenges with outsourcing, backsourcing can be a viable alternative. While outsourcing and related processes have been extensively studied in software engineering, few studies report experiences with backsourcing.

\textbf{Objectives:} We intend to summarize the results of the research literature on the backsourcing of IT, with a focus on software development. By identifying practical relevance experience, we aim to present findings that may help companies considering backsourcing. In addition, we aim to identify gaps in the current research literature and point out areas for future work.

\textbf{Method:} Our systematic literature review (SLR) started with a search for empirical studies on the backsourcing of software development. From each study we identified the contexts in which backsourcing occurs, the factors leading to the decision to backsource, the backsourcing process itself, and the outcomes of backsourcing. We employed inductive coding to extract textual data from the papers identified and qualitative cross-case analysis to synthesize the evidence from backsourcing experiences.

\textbf{Results:} We identified 17 papers that reported 26 cases of backsourcing, six of which were related to software development. The cases came from a variety of contexts. The most common reasons for backsourcing were improving quality, reducing costs, and regaining control of outsourced activities. The backsourcing process can be described as containing five sub-processes: change management, vendor relationship management, competence building, organizational build-up, and transfer of ownership. Furthermore, we identified 14 positive outcomes and nine negative outcomes of backsourcing. Finally, we aggregated the evidence and detailed three relationships of potential use to companies considering backsourcing.

\textbf{Conclusion:} The backsourcing of software is a complex process; its implementation depends on the prior outsourcing relationship and other contextual factors. Our systematic literature review may contribute to a better understanding of this process by identifying its components and their relationships based on the peer-reviewed literature. Our results may also serve as a motivation and baseline for further research on backsourcing and may provide guidelines and process fragments from which practitioners can benefit when they engage in backsourcing.

\end{abstract}

}

\maketitle

\IEEEdisplaynontitleabstractindextext

%
\IEEEpeerreviewmaketitle

\input{1-2-IntroBackground}

\input{3-ResearchMethod}

\input{4-Results}

\input{5-Discussion}

\input{6-Conclusions}

\section*{Contributors}
All authors contributed to conceiving the idea and planning the research. J.M. conducted the search process and collected data from the candidate papers. All authors took part in the study selection process. J.M. and C.L. also carried out data extraction and synthesis of included studies. Finally, all authors discussed the findings and contributed to the final manuscript.

\ifCLASSOPTIONcaptionsoff
 \newpage
\fi



%

\bibliographystyle{IEEEtran}

\appendices

\makeatletter
\renewcommand*{\@biblabel}[1]{[S#1]}
\makeatother

\section{References of Included Studies}
\label{sec:appendix1}
\renewcommand{\section}[2]{}%


\balance

%

\end{document}

%% file: 1-2-IntroBackground.tex
\IEEEraisesectionheading{\section{Introduction}\label{sec:introduction}}

\IEEEPARstart{O}{utsourcing}, i.e., the contracting out of business activities typically performed in-house to third parties, has been reported since the 1970s, and became mainstream in information technology (IT) in the 1990s \cite{davis_it_2006, dibbern_information_2004}. Outsourcing gained considerable media attention due to the benefits advertised by, in particular, a few large, successful initiatives. Cases such as Kodak’s outsourcing of its technology systems in 1989 encouraged other companies to follow the same path \cite{benaroch_should_2010, dibbern_information_2004}. In the following decade, the explosive growth of the internet and major improvements in telecommunications provided an extra boost to this growing trend \cite{dibbern_information_2004, mclaughlin_it_2006}.

Initially, most outsourcing agreements were between companies in the same geographical area. At the beginning of the 2000s, companies began to search for vendors outside their home country, particularly those which could provide additional benefits, such as lower operational costs, access to new markets, or knowledge \cite{davis_it_2006, dibbern_information_2004, solli-saether_stages--growth_2015}. 

As part of the general IT outsourcing trend, companies increasingly started to outsource software development activities as well. In the software engineering research community, this stimulated the emergence of the field of global software engineering \cite{herbsleb_global_2001}, the first conference on which was held in 2006. Research on global software engineering has provided us with an understanding of issues, practices, tools and processes for working in global software engineering projects. Global software engineering, as an empirical research area grounded in real-world challenges, has studied a broad set of issues, the most popular of which have been collaboration and teams, processes and organization, sourcing and supplier management, and success factors \cite{ebert_global_2016}. Since the early 2000s, global software development with various distribution scenarios has become the norm for many companies, either using completely outsourced teams or, increasingly, using companies' own teams spread around the globe in various configurations. 

However, and perhaps not unsurprisingly, not all outsourcing relationships were successful. Many companies ran into problems with their outsourcing initiatives or experienced strategic or environmental shifts \cite{benaroch_should_2010, sparrow_when_2003, wong_bringing_2006} that made the outsourced activity strategically critical to the organization. Expected cost savings were not always realized, often due to unexpected costs in offshore locations \cite{cullen_managing_2006, solli-saether_stages--growth_2015, whitten_adaptability_2010} or coordination and communication costs that proved to be higher than expected. Even when cost savings were achieved, other issues such as poor relationships with the vendor, unsatisfactory quality, and lack of control were reported \cite{brandes_outsourcingsuccess_1997, gottschalk_critical_2005, sparrow_when_2003}. As a result, some companies started reversing their outsourcing decisions, bringing previously outsourced activities back in house, or \emph{backsourcing} \cite{solli-saether_stages--growth_2015, wong_bringing_2006}.

Research on IT outsourcing resulted in a wide range of academic contributions, including investigations into the motivation for outsourcing, expectations about benefits, and the factors leading to a successful outsourcing relationship \cite{gottschalk_critical_2005, lacity_empirical_1998, solli-saether_stages--growth_2015}. Secondary studies have aggregated these findings to understand the outsourcing process, and to synthesize lessons learned from real-world experiences \cite{bergkvist_outsourcing_2008, dibbern_information_2004, khan_barriers_2011, khan_factors_2011}. As the outsourcing phenomenon is multifaceted, new terminologies emerged to help describe the diversity of outsourcing relationships \cite{bergkvist_outsourcing_2008, smite_empirically_2014, solli-saether_stages--growth_2015}.

In contrast to outsourcing, backsourcing has received little attention in the IT research literature \cite{hirschheim_myths_2000, von_bary_westner_information_2018, whitten_bringing_2006}, in which we could identify five studies. As discussed in Section~\ref{sec:relatedwork}, the existing reviews mainly provide insights about the reasons for ending an outsourcing relationship and deciding to bring outsourced IT back into the organization. They provide only a superficial view of the process of backsourcing, and how organizations handle it in practice. Interestingly, we could not find any attempts to aggregate knowledge about backsourcing in the software engineering literature, despite the fact that software development may be either the sole scope, or a significant part of the scope, of IT backsourcing.

Motivated by the lack of in-depth reviews related to the backsourcing of IT, the complete lack of reviews related to backsourcing in software engineering, and the potential contribution both to research and practice, we designed and conducted a systematic literature review to aggregate what is empirically known about backsourcing. Our objective was not limited to investigating the motivation behind the backsourcing decision; rather, we wanted to investigate what is known about the process of backsourcing from decision to completion. In addition to the elements already described in previous reviews, our aim was thus to identify how companies had brought projects back in house, and what the reported outcomes were.

To this end, we searched for and extracted empirical evidence from backsourcing cases reported in the peer-reviewed academic literature. In this paper, we describe the evidence in a narrative format, highlighting the situational context in which the cases were reported. Moreover, to the extent possible, we have tried to explicate relationships between elements such as actions and outcomes. To the best of our knowledge, this paper provides the only in-depth literature review of available empirical studies of backsourcing of software development to date. Thus, the contribution of this paper to the field of software engineering is a) the recognition of the importance of the topic for software engineering research, b) the identification of a lack of in-depth studies of backsourcing in software development, c) a set of empirically-derived insights on backsourcing of value to practitioners who are considering backsourcing as an alternative to their current sourcing strategy, or who are involved in a backsourcing process. Furthermore, we hope to inspire further empirical research that will provide a deeper understanding of the backsourcing process.

The rest of the paper is organized as follows: Section \ref{sec:background} presents background information about backsourcing; Section \ref{sec:research-method} describes our research approach. In Section \ref{sec:results} we report the findings for our research questions. Finally, Section \ref{sec:discussion} compares our results with related work, and discusses the limitations of our study and its implications for research and practice. Finally, Section \ref{sec:conclusions} presents our concluding remarks. 

\section{Background and Related Work}
\label{sec:background}

\subsection{Defining Backsourcing}
\label{sec:backsourcing-definition}
As discussed above, we define \emph{backsourcing} as \emph{the process of bringing previously outsourced activities back in house}. We thus view backsourcing as a process that starts with the decision to backsource and ends when the outsourced activity has been successfully (re-)integrated in the organization. Put another way, backsourcing is the process of moving from outsourcing to insourcing. 

In the literature, the term ``backsourcing'' is used somewhat inconsistently. It has been used as a synonym for reshoring, backshoring, relocating, reverse outsourcing, re-insourcing, or insourcing \cite{bergkvist_outsourcing_2008, nujen_backsourcing_2015, von_bary_how_2018, von_bary_westner_information_2018}. The distinction between these is often not clear, as the contexts in which the terminology is applied differ broadly. Reshoring, backshoring, and relocating are related to moving the outsourced services to a new location, often back to the ``original country'' \cite{nujen_backsourcing_2015}. In contrast, reverse outsourcing and re-insourcing relate to a change in the sourcing relationship \cite{nujen_backsourcing_2015, nujen_managing_2018}, e.g., from outsourcing to insourcing. Insourcing has also been used as a synonym for backsourcing, as reported by a literature review on outsourcing terminology \cite{bergkvist_outsourcing_2008}.

As is clear from our definition, to us the critical factor for backsourcing is the occurrence of a previous outsourcing relationship in which activities such as software development or operations were conducted outside the company's borders. These activities might have been conducted in any geographical location, and as a result of backsourcing may remain in the same location or move to a new location; this is not a defining aspect of backsourcing, but of the new insourcing model with which it ends. Our definition of backsourcing as a process is in line with those of other researchers, e.g., \cite{dibbern_information_2004,hirschheim_myths_2000,lacity_empirical_1998,solli-saether_stages--growth_2015,whitten_bringing_2006,wong_bringing_2006,wong_understanding_2008}.

\subsection{The Emergence of Backsourcing}
\label{sec:backsourcing-in-it}
Backsourcing has been increasingly discussed in the research literature since the 2000s \cite{hirschheim_myths_2000}. In early work, researchers identified four archetypes of insourcing, one of which was carried out as a result of failed outsourcing relationships, i.e., backsourcing. The topic gained some academic interest as reports emerged of companies that had decided to cease outsourcing relationships and were looking for alternative sourcing solutions \cite{dibbern_information_2004, whitten_adaptability_2010}. The most commonly reported post-outsourcing alternatives are re-outsourcing and backsourcing \cite{dibbern_information_2004, whitten_adaptability_2010, whitten_bringing_2006}. Other alternatives include continued outsourcing, switching outsourcing vendors \cite{whitten_adaptability_2010}, and multisourcing relationships \cite{smite_empirically_2014}.

According to the literature, the reasons for termination of an outsourcing relationship are often expectation mismatches, or ``outsourcing expectation gaps'' \cite{lacity_realizing_2007, leyh_information_2018, von_bary_etal_adding_2018, von_bary_westner_information_2018, wong_jaya_drivers_2008}. In particular, gaps occurred in situations where companies rushed into outsourcing agreements expecting major benefits such as reduced costs and improved quality of products and services, but realized later that these expectations were unrealistic. In many cases where the outsourcing effort failed to meet expectations, companies opted to gain full control of the previously outsourced services through backsourcing rather than re-outsource \cite{benaroch_should_2010, bhagwatwar_considerations_2011, nujen_backsourcing_2015}.

The decision to backsource affects both the client and the vendor, and introduces critical issues \cite{nujen_backsourcing_2015, wong_understanding_2008}. One such issue is the transfer of knowledge between the two organizations during the backsourcing process \cite{bhagwatwar_considerations_2011, nujen_backsourcing_2015}. Other tasks that turn backsourcing into a complex process are contract termination and possible negotiation of a new contract to cover the transition period, (re-)building an in-house organization to handle the previously outsourced processes, and ensuring continuity of the previously outsourced services \cite{bhagwatwar_considerations_2011, wong_understanding_2008}.

IT services have become a popular field for outsourcing and, consequently, for backsourcing \cite{kotlarsky_understanding_2012}. The scope of IT backsourcing includes a wide range of services, such as the operation of computer systems (including hardware and software), software development, electronic data processing, and technological support \cite{solli-saether_stages--growth_2015}. In this work, our efforts are focused on understanding backsourcing of software development and its related topics.

\subsection{Existing literature reviews}
\label{sec:relatedwork}

We were able to identify five literature studies related to backsourcing in IT, as shown in Table~\ref{tab:otherstudies}, which lists the reviews and the research questions they address. Notably, there is no review in the field of software engineering, despite the fact that many IT backsourcing projects have a strong software development (SD) component, and that issues related to the reintegration of software engineering in an organization can be among the main challenges in a backsourcing effort. 

\begin{table*}[htbp]
\begin{threeparttable}
\centering
\def\arraystretch{1.2}
\caption{Related literature reviews}
\label{tab:otherstudies}
\begin{tabular}{llllP{.45\textwidth}}
    \toprule
    \textbf{Authors} & \textbf{Type}\tnote{1} & \textbf{Data source} & \textbf{Field} & \textbf{Main topics} \\
    \midrule
    \rowcolor{LightGray}
    Wong and Jaya \cite{wong_jaya_drivers_2008} & NSLR & Press reports & Business & Reasons \\
    Veltri and Saunders \cite{veltri_antecedents_2005, veltri_antecedents_2006} & NSLR & Acad. lit. & - & Reasons \\
    \rowcolor{LightGray}
    Leyh, Schäffer and Nguyen \cite{leyh_information_2018} & SLR & Acad. lit. &- & Reasons and overview of the backsourcing process. \\
    von Bary and Westner \cite{von_bary_westner_information_2018} & SLR/MS & Acad. lit. & Business &  Reasons and implementation success factors. Mapping of topics and research approaches. \\
    \rowcolor{LightGray}
    von Bary et al. \cite{von_researchers_2018} & SLR/MS & Grey lit. & Business & Reasons. Comparison of topics to academic literature.  \\
    \bottomrule
\end{tabular}
\begin{tablenotes}
    \item [1] SLR=Systematic Literature Review; NSLR=Non-systematic Literature Review; MS=Mapping Study
\end{tablenotes}
\end{threeparttable}
\end{table*}

The topic which received the most attention in the reviews was the reason for canceling outsourcing agreements and deciding to backsource. All five reviews included a research question related to this topic, and two of them \cite{veltri_antecedents_2006, wong_jaya_drivers_2008} focused on this question in particular.

Furthermore, the reviews (e.g., \cite{von_bary_westner_information_2018}) do not focus exclusively on empirical experiences, but also include theoretical papers that are not explicitly linked to real-world backsourcing cases. While positioning themselves as systematic literature reviews, two of the papers could better be understood to be mapping studies, outlining what has been studied rather than summarizing the main research findings and what is known about backsourcing \cite{von_bary_westner_information_2018}, \cite{von_researchers_2018}.

%% file: 3-ResearchMethod.tex
\section{Research Method}
\label{sec:research-method}

\subsection{Research Questions}
\label{sec:research-questions}
Our goal was to investigate the phenomenon of backsourcing of software development according to evidence presented in empirical, peer-reviewed studies. We addressed this goal through five research questions:

\begin{enumerate}
  \item[\textbf{RQ1.}] \textit{What is the context of the reported backsourcing instances?} To understand the circumstances of the backsourcing, we extracted characteristics of the business, the previous outsourcing arrangement, and the new organization.
  \item[\textbf{RQ2.}] \textit{Why do companies backsource?} We aim to understand the rationale behind backsourcing, in particular the reported reasons for and against backsourcing. 
  \item[\textbf{RQ3.}] \textit{How do companies backsource?} We want to understand how backsourcing is carried out in practice. What are the elements of the backsourcing process, and how is it performed? 
  \item[\textbf{RQ4.}] \textit{What are the reported outcomes of backsourcing?} We aim to understand the consequences of the backsourcing process. We identified positive and negative outcomes reported in the literature, related practices, and the context in which they occurred.
  \item[\textbf{RQ5.}] \textit{What are the relationships between the context, reasons, processes, and outcomes of backsourcing?} Finally, we intend to deepen our understanding of the backsourcing process beyond what can be achieved by analyzing the themes in isolation. We identified connections between elements we extracted from the literature, and used them to describe potentially interesting relationships. 
\end{enumerate}

\subsection{Search and Selection Process}
\label{sec:search-and-selection}

Our search and selection process is grounded on the guidelines for systematic literature reviews in software engineering \cite{kitchenham_evidence-based_2015}.

\subsubsection{Preliminary Search}
\label{sec:preliminary-search}

To familiarize ourselves with the literature on the topic and to identify relevant databases, we conducted a preliminary search in Google Scholar using a general search string composed of terms related to the phenomenon of interest and research domain. Based on our preliminary search results, we selected five databases: Scopus, ACM Digital Library, Springer Link, IEEE Library, and Web of Science.

We compared our initial search results with the papers included in a related review on the same topic \cite{von_bary_westner_information_2018}, and identified eleven papers missing from the preliminary search that we deemed relevant for our review. By adding two more databases (Science Direct, and EBSCO Host), our search found eight of the missing papers. The three remaining papers were cited by the papers found in our preliminary search, and were thus found by snowballing.

\subsubsection{Search String}
\label{sec:search-string}

Following the preliminary search, we refined our search string using keywords from the most relevant papers. The resulting search string is:

\vspace{.2em}
\noindent
\fbox{\begin{minipage}{0.48\textwidth}
\textit{(backsourcing OR backshoring OR "global resourcing" OR reshoring OR insourcing OR inshoring OR relocating OR "global relocation" OR re-outsourcing OR homeshoring OR "back in-house" OR "fail outsourcing") AND ("software development" OR "software project" OR "software engineering" OR "information technology" OR "information systems" OR "digitalization")}
\end{minipage}}
\vspace{.2em}

We searched the seven databases listed in Table \ref{tab:databases} using the title, abstract, and keyword fields. A full description of the individual search strings used in the different databases is reported in the study protocol\footnote{\label{note:protocol}Available at \url{https://tinyurl.com/ranahvw}}.

\begin{table}[htbp]
\centering
\caption{Databases searched.}
\label{tab:databases}
\begin{tabular}{lrl}
\toprule
\textbf{Database} & \textbf{Results} \\
\midrule
1. Scopus & 278 \\
2. ACM Digital Library & 616 \\
3. Springer Link & 3867 \\
4. IEEE Library & 67 \\
5. Web of Science & 69 \\
6. Science Direct & 44 \\
7. EBSCO Host & 461 \\
\midrule
\textbf{Total} & \textbf{5402} \\
\bottomrule
\end{tabular}
\end{table}

We collected and aggregated the references of the 5402 resulting papers in a single list. We then removed duplicates, incomplete references, and non-papers (see exclusion criteria E4 - E6). This preliminary filtering reduced the number of candidate papers to 3207.

\subsubsection{Study Selection}
\label{sec:study-selection}

The candidate papers were reviewed by two of the authors and selected according to the following criteria:

\textbf{Inclusion Criteria}
\begin{compactenum}[\hspace{18pt}{I}1.]
  \item Empirical studies (e.g., case study, survey, interview) on backsourcing
  \item Grey literature and experience reports on backsourcing cases
\end{compactenum}

\textbf{Exclusion Criteria}
\begin{compactenum}[\hspace{18pt}E1.]
  \item Papers that are not related to the phenomenon of interest, i.e., backsourcing
  \item Papers not in the context of information technology
  \item Secondary studies and theoretical papers
  \item Non-papers, e.g., conference proceedings, lecture notes, and presentations
  \item Non-peer-reviewed papers, e.g., experience reports in the press or on a corporate website
  \item Duplicated papers
\end{compactenum}

Initially, we filtered papers based only on their title and abstract. In the case of papers for which we could not reach a clear decision based only on these criteria, we screened the full text. Two researchers carried out the selection independently, and disagreements were settled by a third researcher. This process resulted in the inclusion of 25 papers.

\subsubsection{Snowballing}
\label{sec:snowballing}

Imprecise terminology resulted in many irrelevant papers that contained the right keywords but did not address the phenomenon we sought to investigate. This created a lot of manual filtering work. At the same time, our preliminary search highlighted a risk that our database searches might have missed relevant papers. To address this problem, we used backward and forward snowballing \cite{wohlin_guidelines_2014} to search for additional studies. 

We first identified a starting set of 25 highly relevant papers through the database search. We had also identified eight secondary studies excluded by criterion E3. The secondary studies are relevant to our research as they potentially identified primary studies we could have missed. The snowballing starting set comprised the 33 papers from these combined sources.

The resulting 1759 papers were added to our selection list. Once again, we removed duplicates and incomplete references, reducing the candidates from the snowballing process to 1202. We applied the selection strategy described in Section \ref{sec:study-selection} to the new candidates, which resulted in the inclusion of four additional papers.

\subsection{Data Extraction}
\label{sec:data-extraction}

In total, we identified 29 papers resulting from our selection strategy: 25 from the database search and four from snowballing. We conducted trial data extraction to confirm whether they contained the data required to answer our research questions. As a result, we excluded 12 papers. The resulting 17 papers are summarized in Table \ref{tab:included-papers}, and a full reference list is provided in Appendix \ref{sec:appendix1}.

\begin{table}[htpb]
\begin{threeparttable}
\def\arraystretch{1.15}
\caption{Summary of the included papers.}
\label{tab:included-papers}
\begin{tabular}{ccllll}
\# & \textbf{Year} & \textbf{Type} & \textbf{Publ. Venue} & \textbf{Field\textsuperscript{1}} & \textbf{Cases} \\ 
\hline
\multicolumn{6}{c}{Case Studies} \\ \hline
\rowcolor{LightGray} 
S2 \cite{wong_bringing_2006}  & 2006 & conf. & ICIS & IS & C23$^{a}$ \\ 
S4 & 2006 & chpt. & & IS & C11-C12$^{b}$ \\ 
\rowcolor{LightGray} 
S5 & 2014 & jour. & ESEJ & SE & C15-C18$^{c}$ \\ 
S6 & 2012 & conf. & ECGSE & SE & C15, C17-C18$^{c}$ \\ 
\rowcolor{LightGray} 
S7 & 2013 & chpt. & & IS & C1-C2 \\ 
S8 & 2011 & conf. & HICSS & IS/IT & C3 \\ 
\rowcolor{LightGray} 
S11 & 2000 & jour. & Comm. ACM & IT & C11-C12$^{b}$ \\ 
S12 & 2008 & conf. & PACIS & IS/IT & C23-C26$^{a}$ \\ 
\rowcolor{LightGray} 
S13 & 2018 & jour. & JIT & IT & C4-C10 \\ 
S14 & 2012 & jour. & JITTC & IT & C13-C14 \\ 
\rowcolor{LightGray} 
S16 & 2016 & mag. & MAGMA & F\&M & C22 \\ 
\hline
\multicolumn{6}{c}{Interview studies} \\ 
\hline
\rowcolor{LightGray} 
S1 & 2018 & jour. & IJISPM & IS & \\ 
S15 & 2018 & jour. & JMTM & TM & C19-C20 \\ 
\rowcolor{LightGray} 
S17 & 2019 & jour. & Israel Affairs & MD & \\ 
\hline
\multicolumn{6}{c}{Survey studies} \\ 
\hline
\rowcolor{LightGray} 
S3 & 2006 & jour. & Dec. Sciences & Bus. & \\ 
S10  & 2016 & conf. & ICEEOT & CNC & \\ 
\hline
\multicolumn{6}{c}{Experience report} \\ 
\hline
\rowcolor{LightGray} 
S9 & 2018 & jour. & JITTC & IT & C21 \\ 
\bottomrule
\end{tabular}
\begin{tablenotes}
\footnotesize
    \item [1] IT: Information Systems; IS: Information Technology; SE: Software Engineering; F\&M: Finance and Management; TM: Technology Management; MD: Multidisciplinary; Bus: Business; CNC: Computer Networks and Communications
\end{tablenotes}
\end{threeparttable}
\end{table}

The backsourcing experiences were mostly described as cases. We identified 26 cases in the included papers. Most of the cases were reported by only one paper, but the cases marked $^{a,b,c}$ in Table \ref{tab:included-papers} were reported by more than one paper. We collected evidence from all the cases, and later merged the redundant information during the data synthesis step.

\begin{compactenum}[a)]
  \item Case C23 was detailed in paper S2, and was also analyzed alongside C24-C26 in the case study S12;
  \item Cases C11 and C12 were reported in the journal paper S11 and also in the book chapter S4; and
  \item Cases C15, C17-C18 were described by the conference paper S6, and also by a more recent journal paper, i.e., S5, complemented by a fourth case. 
\end{compactenum}

We used qualitative data analysis software (NVivo 12) to collect and code the papers and manage an evidence database. The evidence comprises relevant textual information we extracted, a reference to the place in the paper the information was found, and a comment that relates the text to the topics we investigated.

\subsection{Data Synthesis}
\label{sec:data-synthesis}

\begin{figure*}[!hb]
 \centering
 \includegraphics[width=.94\textwidth]{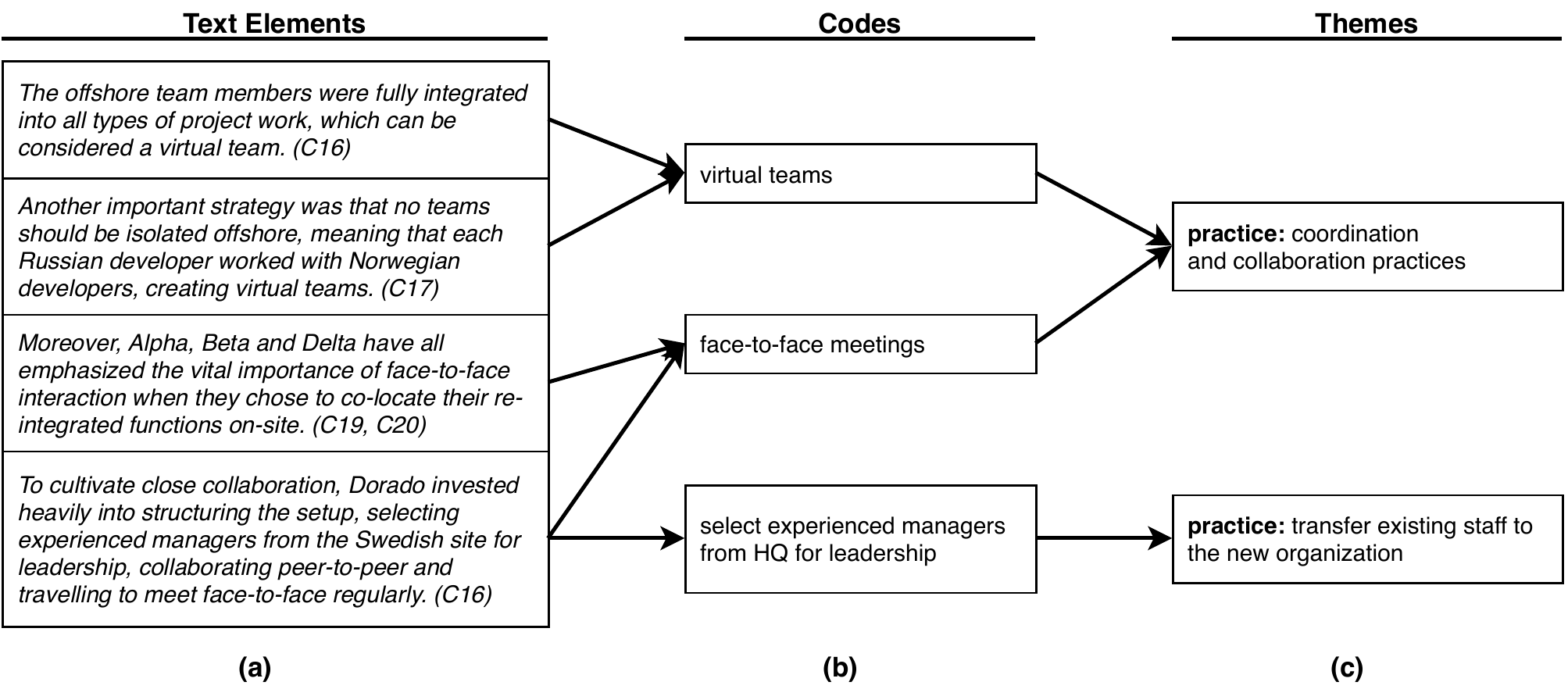}
 \caption{Example of the coding process.}
 \label{fig:example-coding}
\end{figure*}

We used inductive coding and qualitative cross-case analysis \cite{cruzes_case_2015, miles_qualitative_2014, saldana_coding_2015} for data synthesis. The inductive coding process is grounded on codes emerging progressively during the process, instead of starting from a set of pre-defined codes \cite{bailey_qualitative_2003}. Our coding process followed the two-cycle approach \cite{miles_qualitative_2014}, as exemplified in Figure \ref{fig:example-coding}. 

First, we identified textual elements in the papers related to the backsourcing phenomenon (Figure \ref{fig:example-coding}(a)). We then assigned codes to each element that described its content (Figure \ref{fig:example-coding}(b)). A textual element can be associated with multiple codes. Further, we checked codes across the whole data set to identify similarities. We merged codes with the same meaning, and aggregated similar codes into higher-level codes called themes (Figure \ref{fig:example-coding}(c)). We often needed to go back to the previous step and revise the codes, changing their description to better fit the ``overall picture'' provided by the themes.

We carried out this process several times, deriving higher-level themes with each iteration. We stopped aggregating codes when high-level themes associated with the research questions emerged, i.e., \textit{contextual information, reason for backsourcing, elements of the backsourcing process,} and \textit{outcomes of backsourcing}. We conducted further revisions to prune and organize the themes into meaningful structures which we could use to interpret the findings.

To present the findings, we employed matrix display and narrative description. The matrices tabulated and organized the data – codes and themes – for interpretation. The narrative descriptions portrayed the findings according to the contextual situations reported in the papers, and allowed us to compare situations across cases, highlighting similarities and differences \cite{cruzes_case_2015, miles_qualitative_2014}. Matrix display was used to address RQ1 to RQ4, and narrative description was used to describe the themes related to RQ2 to RQ5.

Finally, we established connections between the themes using short comments to explain such a relationship (e.g., supports, mitigates, incurs). We used relationships to create networks that better described a backsourcing event in the light of a situation or context \cite{miles_qualitative_2014} and helped us to address research question RQ5. Our approach to investigating relationships was grounded on axial coding. First, we identified themes containing a higher number of relationships and chose it as a central entity for the relationship network. We then draped around the central entity the themes that were linked to it, drawing lines to represent the relationships. Successive iterations of this step expanded the network from its core to its edges.

\subsection{Threats to Validity}
\label{sec:validity-threats}

This section discusses the threats to the validity of our study and the actions we took to mitigate them, according to the categories of Ampatzoglou et al. \cite{ampatzoglou_identifying_2019}:

\textbf{Study selection validity.} Completeness is a critical issue for SLRs \cite{kitchenham_evidence-based_2015}. Our mixed search approach considered seven electronic databases and further snowballing interactions, aiming to identify all the relevant papers. We employed validation steps to reduce the risk of missing relevant papers, including a preliminary search and a comparison with related work. Another threat related to completeness regards the precision of the search string due to multiple meanings and homonym terms. Unreliable terms, e.g., ``resourcing'', broadly increased search results with too many irrelevant papers. Through multiple iterations, we refined our search string to find relevant studies.

Prior to the selection, we piloted a subset of the papers and revised the inclusion and exclusion criteria, aiming for a common understanding. To further reduce the likelihood of excluding relevant papers, two reviewers carried out the selection process independently and a third researcher mediated disagreement. We extended our selection process to cover papers in languages other than English, but only a Norwegian paper was included. We found candidate papers in other languages (e.g., German, French, and Chinese), but they were later excluded, as they did not provide evidence to address our research questions.

\textbf{Data validity.} Our resulting data set comprises 17 papers that investigate different aspects of the backsourcing phenomenon. This data set is probably not a significant sample of the backsourcing cases. In particular, we did not observe saturation in our coding process that provided enough evidence to depict the backsourcing phenomenon. The scarce evidence about the phenomenon of interest limited the conclusions we could draw from the findings. In addition, the high heterogeneity of the included papers poses a challenge, as evidence gathered from a given primary paper was often not reinforced by other papers \cite{miles_qualitative_2014}.

The lack of a quality assessment of primary studies is a deviation from the SLR process \cite{kitchenham_evidence-based_2015}. As a result, we could not compare the evidence from multiple primary sources based on a set of quality criteria. Aiming to compare cross-case findings, we used the frequency with which the themes occurred, and the source of evidence, e.g., personal observation, reported by a participant.

We also acknowledge that our inductive coding was influenced by our ability to reflect upon the data and biases due to our experiences \cite{ampatzoglou_identifying_2019, bailey_qualitative_2003}. We tried to mitigate issues of neutrality and impartiality via a team-based data extraction/synthesis using NVivo server as the shared database. One researcher carried out the data extraction and checked with the other two authors. Recurrent discussions during the entire process provided opportunities to identify and correct researcher biases. 

\textbf{Research process validity.} Repeatability is another essential aspect of SLR validity\cite{kitchenham_repeatability_2011}. To achieve this, we followed the guidelines for systematic literature reviews in SE \cite{kitchenham_evidence-based_2015} and provided a detailed SLR protocol\footnoteref{note:protocol}. Any decision points and deviations from the proposed process were reported. To strengthen the reliability of our qualitative data extraction and synthesis, we have also made our nVivo data set available upon request.

Another potential area of bias is related to whether the questions address the main goal of the research \cite{kitchenham_evidence-based_2015}. Our main goal was broken down into five research questions: RQ1 is a supporting tool that helped us to contextualize the findings from the other questions; RQ2, RQ3, and RQ4 were addressed independently, and we later compared our findings with those of related work (see Section \ref{sec:discussion-comparison}); RQ5 is a novel contribution that emerged from our iterative data synthesis process.

%% file: 4-Results.tex
\section{Results}
\label{sec:results}

We identified 26 case studies of backsourcing in 13 of the included papers. Two interview studies and two surveys, all with practitioners experienced in outsourcing and backsourcing as subjects, provided additional information. All included papers reported studies of practical, real-world backsourcing experiences.

\subsection{Overview of the Reported Backsourcing Cases}
\label{sec:results-overview}

Table \ref{tab:context} presents the organizational context of the 26 identified backsourcing cases. A diverse set of business sectors was observed; frequently, those reported included software products and services (cases C15-C18, and C22) and financial services (cases C2, and C4-C6). A range of different activities has been subject to outsourcing, such as application hosting (cases C1, C2, C4-C6, C10, and C14), software development (C13, C15-C18, and C21), data center management (C8, C9, and C14) and server management services (C7). Our dataset also includes a few cases of backsourcing of undefined IS/IT services (cases C3, C11, and C22), and  six cases that did not detail the backsourced services at all. The cases that explicitly concerned software development are marked in bold in the table. 

\afterpage{
    \clearpage
    \thispagestyle{empty}
    \begin{landscape}

    \begin{table}[htbp]
    \def\arraystretch{1.67}
    \caption{Organizational context of the cases.} 
    \label{tab:context}
    \begin{tabular}{lP{.25\textwidth}P{.45\textwidth}P{.15\textwidth}P{.15\textwidth}P{.15\textwidth}}
    \toprule
    \textbf{Case} & \textbf{Business sector} & \textbf{Backsourced services} & \textbf{Company size / New organization size} & \textbf{Pre-backsourcing sourcing model} & \textbf{Post backsourcing sourcing model} \\ 
    \midrule
    \rowcolor{LightGray} 
    C1 (S7) & Food manufacturing & Application and infrastructure hosting and management & - & outsourcing & outsourcing \\ 
    C2 (S7) & Banking and finance & Application hosting & Large (39k) & outsourcing & outsourcing  \\ 
    \rowcolor{LightGray}
    C3 (S8) & - & IS/IT services (not detailed) & Large (150k) & outsourcing & onshore insourcing \\ 
    C4 (S13) & Banking and finance & Application hosting & Large (140k) & offshore outsourcing & onshore insourcing \\ 
    \rowcolor{LightGray}
    C5 (S13) & «same company as C4» & Application hosting and management & «same as C4» & onshore outsourcing & onshore insourcing \\ 
    C6 (S13) & «same company as C4» & Application hosting & «same as C4» & outsourcing &  out- and insourcing \\ 
    \rowcolor{LightGray}
    C7 (S13) & Pharmaceutical, biological and consumer healthcare & Server management and monitoring & Large (99k) & outsourcing & onshore insourcing \\ 
    C8 (S13) & Insurance & Datacenter and operations & Large (20k) & offshore outsourcing & onshore and offshore insourcing \\ 
    \rowcolor{LightGray}
    C9 (S13) & Automotive & Datacenter and data management & Large (25k) & offshore outsourcing & onshore insourcing \\ 
    C10 (S13) & Chemical manufacturing, transport, and logistics & Application hosting and management & Large (8k) & outsourcing & onshore insourcing \\ 
    \rowcolor{LightGray}
    C11 (S4, S11) & Chemical manufacturing & IT services (not detailed) & Small (40) & outsourcing & onshore insourcing \\ 
    C12 (S4, S11) & Chemical manufacturing & IT services (not detailed) & Large (1k) & outsourcing & onshore insourcing \\ 
    \rowcolor{LightGray}
    \textbf{C13 (S14)} & \textbf{IT services provider} & \textbf{Software development }& \textbf{Large (1.5k)} & \textbf{offshore outsourcing} & \textbf{onshore insourcing} \\ 
    C14 (S14) & CD \& DVD manufacturing & SAP application and database hosting & Large (8k) & outsourcing & onshore insourcing \\ 
    \rowcolor{LightGray}
    \textbf{C15 (S5, S6)} & \textbf{Software-intense services} & \textbf{Software development} & \textbf{- / Small (40)} & \textbf{offshore outsourcing} & \textbf{offshore insourcing} \\ 
    \textbf{C16 (S5)} & \textbf{Software product} & \textbf{Software development} & \textbf{- / Medium (100)} & \textbf{offshore outsourcing} & \textbf{out- and insourcing} \\ 
    \rowcolor{LightGray}
    \textbf{C17 (S5, S6)} & \textbf{Software product} & \textbf{Software development} &\textbf{ Large (260$\sim$270) / Small (60)} & \textbf{offshore outsourcing} & \textbf{offshore insourcing} \\ 
    \textbf{C18 (S5, S6)} & \textbf{Software product} & \textbf{Software development and maintenance} & \textbf{- / Medium (100)} & \textbf{offshore outsourcing} & \textbf{offshore insourcing} \\ 
    \rowcolor{LightGray}
    C19 (S15) & Telecom & - & - & outsourcing & onshore insourcing \\ 
    C20 (S15) & Telecom & - & Large (21k) / Large (300) & outsourcing & onshore insourcing \\ 
    \rowcolor{LightGray}
    \textbf{C21 (S9)} & \textbf{Government} & \textbf{Software development and operations} & \textbf{-} & \textbf{outsourcing} & \textbf{onshore insourcing} \\ 
    C22 (S16) & Software provider & - & - & offshore outsourcing & onshore insourcing \\ 
    \rowcolor{LightGray}
    C23 (S2, S12) & Service & IT services (not detailed) & Large (10k) & outsourcing & onshore insourcing \\ 
    C24 (S12) & Higher education (private) & - & - & outsourcing & onshore insourcing \\ 
    \rowcolor{LightGray}
    C25 (S12) & Government & - & - & outsourcing & onshore insourcing \\ 
    C26 (S12) & Consumer goods & - & - & outsourcing & onshore insourcing \\
    \bottomrule
    \end{tabular}
    \end{table}

    \end{landscape}
    \clearpage
}

All backsourcing cases were from different companies, except for C4-C6, which were from the same financial services company, each of whose cases described the backsourcing of the application hosting of a different service.

The table also shows the size of the company and the size of the new organization created to handle the previously outsourced activities. Using the EU classification\footnote{\url{https://ec.europa.eu/regional_policy/sources/conferences/state-aid/sme/smedefinitionguide_en.pdf}} for companies, 15 cases were from large companies with more than 250 employees, and one case was from a small company (C11, with 40 employees). The size of the new organization designed to handle the backsourced activities was reported by only five cases (C15-C18, and C20): this ranged from small (less than 50 employees) to large (250+ employees).

We classified the sourcing models according to their relationship and geographical location \cite{aagerfalk2008outsourcing,smite_empirically_2014}. For the relationship, outsourcing means that an outside vendor provided the service, while insourcing means that the service is produced within the company. The purpose of backsourcing is to transition from outsourcing to insourcing. However, we identified two cases (C6 and C16) which ended in a mixed-sourcing strategy. Our dataset also contains two cases in which backsourcing was considered but rejected (C1 and C2). Although these cases do not describe a backsourcing process, we included them because they provide meaningful insights about the reasons for and against backsourcing.

With respect to the geographical location, we identified backsourcing cases where the location of the new organization was either onshore, i.e., in the same country as the parent organization, or offshore, i.e., in another country. In their outsourcing agreements, most of the companies had used offshore vendors; an exception was case C5, which used a local vendor. After the termination of outsourcing, the most common strategy was onshore insourcing, i.e., bringing the service geographically close to the outsourcing organization. Four cases (C8, C15, C17, and C18) transferred from offshore outsourcing to ``offshore insourcing'', using offshore subsidiaries to take over the outsourced activities. We also identified one case (C8) of a mixed strategy whereby some services were brought back in-house, while others were transferred to different offshore locations. 

The overview in Table \ref{tab:context} shows the set of different contexts in which the backsourcing cases occurred. Only six of the cases are related to software development, and four of these are from the same study (papers S5 and S6). The organization size in these cases varied from small, with 40 employees, to large, with more than 1500 employees. Only C13 reported \textit{onshore insourcing} as the post-backsourcing model. Cases C15-C18 opted for backsourcing to an offshore subsidiary of the company, and C21 failed in its backsourcing process, reverting to the original outsourcing agreement.

\subsection{Reasons for and against Backsourcing}
\label{sec:results-reasons}

Decisions for or against backsourcing were primarily made by top management and IT leaders, often taking into account the perceptions (and complaints) of operational staff. In particular, developers' opinions have been used to support decisions in cases of dissatisfaction with vendor quality.

All but one of the included papers (S4) discuss the reasons for the backsourcing decisions. The papers did not share a standard classification of reasons. We therefore identified text elements related to reasons and grouped them into categories with similar meanings, as explained in Section \ref{sec:data-synthesis}. For example, the high-level reason \textit{quality problems} is made up of low-level codes such as the quality of services provided and the quality of software products delivered. By this method we established seven reasons for backsourcing and three against, as shown in Table~\ref{tab:reasons}.

We identified reasons against backsourcing in papers that discuss a decision between backsourcing and other sourcing options, e.g., switching vendors or continuing with outsourcing. In this context, the reasons against backsourcing are points in favor of another sourcing option. In most cases, companies adopted backsourcing in spite of the reasons against it. In only two cases (C1 and C2, in paper S7) did they opt to switch vendors instead. 

The first six rows represent the cases in which the backsourcing of software development activities was explicitly discussed, and the other 17 rows represent cases of backsourcing of other IT-related services, or cases in which the backsourcing service was not reported (C19, C20, C24-C26, see Table~\ref{tab:context}). Looking at the table, we can make some general observations before embarking on a more detailed discussion of the individual reasons for backsourcing. 

The most commonly reported reasons for backsourcing in the included papers were \textit{quality problems} and \textit{high costs}. These two reasons, together with \textit{lack of control}, have been discussed under the heading of ``outsourcing expectation gaps'' in both outsourcing and backsourcing studies \cite{lacity_realizing_2007, leyh_information_2018, von_bary_etal_adding_2018, von_bary_westner_information_2018, wong_jaya_drivers_2008}. This gap refers to situations where companies rushed into outsourcing agreements expecting to save costs and improve quality while retaining control of the outsourced product or service. However, their experience shows that these expectations were unrealistic and were not realized. This in turn led clients to reconsider their sourcing strategies, sometimes opting for backsourcing \cite{wong_jaya_drivers_2008}.

The number of reasons leading to backsourcing reported in the individual cases ranges from one to six. From the papers reporting multiple reasons, the ones that often co-occurred were \textit{a poor client-vendor relationship} and \textit{quality problems} (S1, S3, C16-C18, C19,-C20). \textit{Quality problems} were often mentioned alongside other reasons. 

Another pattern we identified was the co-occurrence of \textit{a poor client-vendor relationship} as a reason to backsource, and \textit{lack of internal capabilities} as a reason not to (S1, C15, C18-C19). Among the cases that reported multiple reasons against a backsourcing decision, we also found that \textit{lack of internal capabilities} and \textit{dependency of the vendor} co-occurred in three cases (S1, C19-C20). 

It is important to note that S1 and C19 displayed all the patterns mentioned above. S1 is an interview study that gathered evidence from 16 participants working as consultants or independent advisors in backsourcing. The paper reported six reasons, i.e., four in favor of the backsourcing decision and two against. C19 is a case of backsourcing from the telecom industry that presented a problematic situation. Cost savings were reported as the official reason for backsourcing, but the evidence from interviews with the management revealed other issues with the outsource vendor, including competence gaps.

Comparing the software development cases with the others, we note that they share many of the same reasons for backsourcing. However, the reason most commonly reported in such cases was a \textit{poor client-vendor relationship} instead of the reasons associated with expectation gaps. Interestingly, just two out of the three reasons associated with the gaps were reported; \textit{high costs} were not reported by a single software development case, nor were \textit{changes in strategy or management}, \textit{outdated technology}, or \textit{new regulations}. Furthermore, no software development case reported \textit{dependency on the vendor} as a factor that would have detracted from the decision to backsource.

\begin{table}[htbp]
\def\arraystretch{1.2}
\caption{Reasons for and against backsourcing}
\label{tab:reasons}
\begin{tabular}{l|cccccccc|ccc}
\textbf{Case} & \rot{\textbf{Quality problems}} & \rot{\textbf{High costs}} & \rot{\textbf{Lack of control}} & \rot{\textbf{Poor client-vendor relationship}} & \rot{\textbf{Vendor competence issues}} & \rot{\textbf{Changes in strategy or management}} & \rot{\textbf{Outdated technology}} & \rot{\textbf{New regulations}} & \rot{\textbf{Lack of internal capabilities}} & \rot{\textbf{High switching costs}} & \rot{\textbf{Dependency on the vendor}} \\
\hline
 & \multicolumn{8}{c|}{\textit{Reasons for}} & \multicolumn{3}{c}{\textit{... against}} \\
 \hline
\multicolumn{12}{c}{\textbf{Software development cases}} \\
\hline
\rowcolor{LightGray}
C13 & & & $\bullet$ & & & & & & & $\bullet$ & \\
C15 & & & & $\bullet$ & & & & & $\bullet$ & & \\
\rowcolor{LightGray}
C16 & $\bullet$ & & & $\bullet$ & & & & & & & \\
C17 & $\bullet$ & & & $\bullet$ & $\bullet$ & & & & & $\bullet$ & \\
\rowcolor{LightGray}
C18 & $\bullet$ & & & $\bullet$ & $\bullet$ & & & & $\bullet$ & & \\
C21 & & & $\bullet$ & & & & & & & & \\
\hline
\multicolumn{12}{c}{\textbf{Other IT-related cases}} \\ 
\hline
\rowcolor{LightGray}
C1 & & $\bullet$ & & & & & & & & $\bullet$ & $\bullet$ \\
C2 & & $\bullet$ & & & & & & & & $\bullet$ & \\
\rowcolor{LightGray}
C3 & & $\bullet$ & $\bullet$ & & & $\bullet$ & & & & & \\
C11 & $\bullet$ & $\bullet$ & & & $\bullet$ & & & & & & \\
\rowcolor{LightGray}
C12 & & $\bullet$ & & & $\bullet$ & & & & & & \\
C14 & & $\bullet$ & $\bullet$ & & & & & & & & \\
\rowcolor{LightGray}
C19 & $\bullet$ & & & $\bullet$ & $\bullet$ & & & & $\bullet$ & & $\bullet$ \\
C20 & & & & & $\bullet$ & & & & $\bullet$ & & $\bullet$ \\
\rowcolor{LightGray}
C22 & $\bullet$ & & $\bullet$ & $\bullet$ & & & & & & & \\
C23 & $\bullet$ & $\bullet$ & $\bullet$ & & & $\bullet$ & $\bullet$ & & & & \\
\rowcolor{LightGray}
C24 & $\bullet$ & & $\bullet$ & & & $\bullet$ & & & & & \\
C25 & & & & & $\bullet$ & $\bullet$ & $\bullet$ & & & & \\
\rowcolor{LightGray}
C26 & & & & & & $\bullet$ & & & & & \\
S1 & $\bullet$ & $\bullet$ & & $\bullet$ & & & & $\bullet$ & $\bullet$ & & $\bullet$ \\
\rowcolor{LightGray}
S3 & $\bullet$ & $\bullet$ & & $\bullet$ & & & & & & & \\
S10 & $\bullet$ & $\bullet$ & & & $\bullet$ & $\bullet$ & & & & & \\
\rowcolor{LightGray}
S17 & $\bullet$ & $\bullet$ & $\bullet$ & & & & & & & & \\
\hline
\textbf{Total} & \textbf{12} & \textbf{11} & \textbf{8} & \textbf{8} & \textbf{8} & \textbf{6} & \textbf{2} & \textbf{1} & \textbf{5} & \textbf{4} & \textbf{4}
\end{tabular}
\end{table}

\subsubsection{Reasons for Backsourcing}
\label{sec:results-reasons-for}
As can be seen in Table~\ref{tab:reasons}, we identified the following reasons for backsourcing: 

\textbf{Quality problems.} 12 cases mentioned quality problems or the need to improve quality as a reason for backsourcing, making this the most cited reason in our data. Three of the software development cases mentioned quality issues, making this the second most cited reason in that category. Quality issues mentioned included poor quality of the delivered product (C4-C10, C16-C18, C19, and C22), poor service provided (C16, C22, C23, S3, and S17), end-customer dissatisfaction (C11, C24, and S1), and delayed deliveries (C19 and C22). In many cases an attempt to solve quality issues in the outsourcing agreement put a strain on the relationship, perhaps exacerbating existing relationship problems, as a team leader from C18 commented: \textit{``Our people felt like they were spending basically all their time writing work orders and writing code for these guys through [email].''} A survey-based study (S3) investigated whether the companies chose to switch vendors or to backsource. The participants represented a range of companies from sectors such as manufacturing, education, healthcare, and public administration. Out of 160 respondents, 70 kept their outsourcing agreement, 54 opted for backsourcing, and 36 switched vendors. The decisive argument for choosing backsourcing rather than switching vendors was poor quality of products and services. In C17, the need to improve quality played a vital role in favor of the backsourcing decision.
  
\textbf{High costs.} The second most cited reason, reported in 11 cases, was high costs. Cases C3, C11, C12, and C14 and two other studies (S1 and S17) reported that outsourcing could be more expensive than expected due, e.g., to poorly-negotiated contracts and unexpected costs. In C3, rising outsourcing costs also exposed quality issues, and the company felt they were not receiving ``value for money''. In C23, S1, and S3, backsourcing was perceived to be cheaper than continued outsourcing (C23, S1, and S3) due to the high expected costs of solving quality issues. C1, C14, and S13 reported that backsourcing was cheaper than switching vendors due to the extra costs incurred in finding another vendor, and the potential savings from setting up an in-house environment. In C2, C14, and S3, expectations of lower in-house maintenance and operational costs motivated the backsourcing decision. In C3, cost savings were the official reason for backsourcing, but the researchers raised the question of whether the real reason was the CEO's perception of a bad outsourcing agreement. If this was the case, high costs were being used as an excuse to terminate the outsourcing agreement. 
  
\textbf{Lack of control.} Eight studies reported that companies experienced a lack of control over their projects or services due to outsourcing. Loss of control of companies' own services (C14, C21), lack of control over the vendor (S17) and a wish for more flexibility in the management of the sourced project (C13, C14, C23, and C24) were reported as reasons to backsource. In cases C3 and C24, the outsourced service gained vital importance for the company due to business changes, justifying the backsourcing decision.
  
\textbf{Poor client-vendor relationship.} Degradation of outsourcing relationships leading to a backsourcing decision was reported in eight studies, four of which concerned software development activities. In four cases (C16, C17, C18, and C22), poor communication and collaboration issues were reported causes of relationship degradation. Other factors included internal staff dissatisfaction (C16 and C17), lack of trust in the vendor (C22 and S3), conflicts over product ownership (C17), and misalignment between the client and vendor organizations' ways of working (S1). In cases C15, C16, C18, and C19, the deterioration of the outsourcing relationship was driven by a perception of poor service quality. One manager in C16 commented, \textit{``What we found out with [vendor's name] was that, you know, the maintenance team, took only the easy bugs, and they were measured on the number of solved bugs. They took the easy ones, not the critical ones.''}
  
\textbf{Vendor competence issues.} Lack of vendor competence was reported in C17, C18, C20, and C25. Building vendor competence required a great deal of effort and resources from the client in C11 and C12, and the resulting knowledge gained by the vendor was subsequently not used in the best interests of the client. A manager in C11 recalled, \textit{``I think you find with outsourcing that any innovation in technology comes from your own people, (…) But basically the [outsourcing vendors] just crank it. (…) You pay for them to learn your business, then they move those people to court other companies in your industry. They transfer skills to get new business, now the learning curve is yours to pay for again.''} In C11, the vendor refused to introduce new technologies and siphoned talents to other customers, while as in C12, the client provided the vendor with its own technical staff. In C20, the vendor's expertise ceased to be a market differentiator, prompting the company to start discussing backsourcing.
  
\textbf{Changes in strategy or management.} Changes in core competency (S10), changes of strategic direction (C23, C24), and changes in management (C23, C25, and C26) were reported as impetuses to review a company's sourcing strategy. In C24, a new business partnership brought an opportunity to review existing outsourcing agreements. In cases C3 and C23, decision-makers' personal beliefs and attitudes were essential drivers of the backsourcing decisions. 
  
\textbf{Outdated technology.} In C23 and C25, the vendor employed an outdated technology, causing asymmetries with the client's organization that motivated the decision to backsource. In C25, an interviewee commented \textit{``The equipment [the vendor used] was relatively outdated, older. A lot of the systems were truly dispersed systems (…) with little capability of acting and interacting with other processes.''}
  
\textbf{New regulations.} Five out of 12 participants of an interview study (S1) mentioned how compliance with new regulations (e.g., data privacy laws, and bank regulations and standards) forced the client to bring services back in-house. The authors of S1 noted that this reason appeared only in the most recent interviews, due to the recent introduction of stricter standards by regulatory bodies in Europe. 

\subsubsection{Reasons against Backsourcing}
\label{sec:results-reasons-against}
As shown in Table \ref{tab:reasons}, we identified the following three reasons against backsourcing:

\textbf{Lack of internal capabilities.} Four cases (C15, C18-C20) and one interview study (S1) reported lack of internal capabilities as a hindrance to backsourcing. C19 and C20 pointed out how longer outsourcing relationships exhausted the company's capabilities to transfer services back. S1 described factors such as missing staff, lack of technical knowledge, and lack of support from the vendor that prevented companies from backsourcing. Such reasons led the company in C1 to reject the backsourcing decision.
  
\textbf{High switching costs.} Changing the sourcing strategy, either by backsourcing or switching vendors, was reported as a high-cost process by C1, C2, and C17. Case C13 reported discussions about higher costs of operational activities after bringing services back in-house. In C1, \textit{high switching costs} were reported as one of the reasons for opting for a switch of vendors over backsourcing.
  
\textbf{Dependency of the vendor.} In C21, operational dependencies on the vendor locked the client into an undesired outsourcing agreement. Similar dependencies were reported in C19 and S1, related to accessing expert knowledge, and to the lack of support for the transfer back, respectively. The issue was aggravated in C19, as the client shared the knowledge with the vendor in the first place, as one participant recalled: \textit{``First we had to share our knowledge on how to produce telecom products […] now they have this expertise and are competing with us. They hold a trump card, knowing that they are now the only ones who have this expertise […] To resolve this, we now have to pay them a huge amount in order to be able to take back this part of the product in-house […]''}

\subsection{Backsourcing as a Process}
\label{sec:results-elements}

One of our main goals was to establish what is known about the process of backsourcing. However, based on the current literature, this turned out to be rather challenging. Two papers (S2 and S8) described the backsourcing process as consisting of sequential stages. The processes they describe have few similarities in terms of terminology or the logic behind the formation of the process stages. Most papers simply described what companies had done as a connected sequence of events, sometimes including some rationale. Therefore, instead of trying to force the reported findings into a preconceived model, we employed selective coding and successive refinement iterations, identifying elements of the descriptions and their relationships. We identified the following set of elements of the backsourcing process:

\textbf{Sub-processes:} Fragments of the overall backsourcing process that describe essential things taking place (see Sections \ref{sec:element1} to \ref{sec:element4}). Although the sub-processes we identified can seem similar to phases or stages of the conceptual models, they are different in the sense that they were not necessarily carried out sequentially. They might have been partially ordered, and conducted simultaneously, and iteratively. 

\textbf{Categories:} Segments of a sub-process intended to group other backsourcing elements (activities, artifacts, attributes and practices) that share a similar goal.

\textbf{Activity:} A piece of work undertaken as part of the backsourcing process or its sub-processes. In our coding process, activities are fine-level entities that describe actions that aim at producing a desired outcome. Activities differ from sub-processes and their categories, as those merely group multiple activities (and other elements) that have a shared goal. Activities are listed in relation to sub-processes in Sections \ref{sec:element1} to \ref{sec:element4}. 

\textbf{Artifacts:} An object or piece of information observed during the backsourcing process, which could affect or be affected by activities or sub-processes. In our study, artifacts mostly provide information support about the process or its elements. As an example of an artifact, a \textit{transition plan} helps to detail the \textit{knowledge transfer} activity.

\textbf{Attributes:} Qualities or features that characterize a sub-process, activity, or artifact. We did not try to gather an extensive list of attributes, but instead tried to identify essential characteristics that, according to the cases, could affect the backsourcing process. As an example, \textit{implicit} and \textit{explicit knowledge-types} are discussed according to different activities required for building knowledge in the new in-house organization.

\textbf{Practices:} Practices offer a certain way of performing an activity, or they support activities through the application of a method or approach. The cases we identified in our study often described different ways of enacting the same activity. As an example, \textit{early termination} and \textit{delay termination} practices are different ways of  handling the termination of an existing outsourcing agreement. We designated practices with an identifier (e.g., P1) so that we could associate them with the outcomes to which they contribute (see Section \ref{sec:results-outcomes}).

\begin{table*}[!hb]
\def\arraystretch{1.1}
\caption{Change management} 
\label{tab:element1}
\begin{tabular}{P{.1\textwidth}P{0.12\textwidth}P{0.05\textwidth}P{0.63\textwidth}}
\toprule
\textbf{Category} & \textbf{Element} & \textbf{Type} & \textbf{Observations} \\
\midrule
Planning & Plan backsourcing efforts & activity & In C11, C21, and C23, a backsourcing plan outlined the tasks and efforts required for competence build-up and organizational build-up (see planning artifacts in Tables \ref{tab:element1b} and \ref{tab:element2}). In C23, the plan helped to communicate updates in the backsourcing process to top business management and other organizations inside the company. \\ \addlinespace
 & Backsourcing plan & artifact & The plan in C21 included a timeline, activities, and related practices, and the rationale for the adoption of such practices. No case reported the people or roles involved in the process, nor evaluation steps or criteria for assessing the outcome of backsourcing. \\ \addlinespace
 & Backsourcing scope & artifact & Defines the scope of the backsourcing, including, but not limited to, the outsourced service or product. S1 describes careful consideration about the scope of outsourced resources and services to transfer back. In cases C13-C15, and C20, the client opted to transfer back all knowledge previously outsourced, whilst C4-C6 reported a partial backsourcing scope. \\ \addlinespace
 & New sourcing location & attribute & Describes the sourcing location for the new organization. Companies need to consider relocation options (S1 and C23). As well as onshore, nearshore (S1) or offshore subsidiary (C15-C18) were also considered. In some cases, offshore alternatives offered advantages: lower recruiting costs (S1 and C18), access to highly skilled personnel (C16 and C18), and proximity to the customer market (C16 and C17). Offshore challenges included administrative overheads (C15 and C16), alignment of ways of working (C15), and ensuring a shared culture (C17). \\ \addlinespace
 & Select what to backsource (P1) & practice & Selective transfer approaches were employed to select what to backsource in cases C3 and C23. The organization in C23 evaluated components by a cost-benefit analysis comparing bringing back in-house versus re-outsourcing. Also in S1, interviewees stated that they first backsourced all previously outsourced services, then they selectively outsourced some of them afterwards. \\ \addlinespace 
& Form a planning team (P2) & practice & In C21, a planning team laid out the steps for backsourcing. The team carried out a pilot backsourcing process, identifying and documenting issues; they also elaborated a transition plan (see Table \ref{tab:element2}) to guide the knowledge transfer. \\ \addlinespace
 & Establish a contingency plan (P3) & practice & C3 and C21 reported a need to establish potential risks and mitigations. The difficulty of managing unanticipated obstacles was also highlighted in C23. In C3, the contingency plan mitigated the effects of a complete relationship breakdown by speeding up knowledge transfer (see Section \ref{sec:element2}). \\ 
\midrule
Internal Communication & Communicate the backsourcing decision & activity & Internal communication was the  starting point for the backsourcing processes in S1, C11, C12, and C23. Reasons leading to the backsourcing decision were made explicit and communicated to the in-house organization (S1, C11, and C12), motivating the backsourcing process. \\ \addlinespace 
 & Sell the backsourcing case internally (P4) & practice & In C11 and C23, this practice was used to inform top management and other organizations inside the company about the backsourcing plan. In C23, further communication provided updates of the backsourcing process and detailed changes affecting other internal organizations.\\ 
\midrule
Post-backsourcing & Monitor outcomes of backsourcing & activity & Case C23 reported how the organization specified performance measures to gauge the success of backsourcing. Unfortunately, the study does not describe the measurement approach used; rather, the perceptions of stakeholders (mostly managers) are reported. Similar performance assessments were also reported in C3 and C11-C14. \\ 
\addlinespace
 & Post-backsourcing expansion & activity & A follow-up activity reported by a few cases (C15-C18) of offshore insourcing. Companies expanded their in-house services even further after backsourcing, by expanding their offshore organizations (C16 and C18), or via new business partnerships (C15-C18, S1). \\
\bottomrule
\end{tabular}
\end{table*}

The elements of backsourcing we identified helped us outline the backsourcing process as observed in the cases of interest. These elements do not provide a complete representation of the process, as they are limited by the context of reported cases and the quality of the information in the articles. Fleshing out the elements identified here into a fully-fledged description of a potentially ideal backsourcing process has been left for future work.

As a result of our coding process, we identified five sub-processes of backsourcing: \textit{change management, vendor relationship management, competence building, organizational build-up,} and \textit{transfer of ownership.} Each sub-process describes a main theme of the backsourcing process. We describe each one according to its context in the papers, and list the related elements (i.e., activities, artifacts, attributes, and practices) in Tables \ref{tab:element1} to \ref{tab:element4}. Each table represents one sub-process and is further divided into activities. In this section we present the sub-processes identified, and in Section \ref{sec:results-relationship} we discuss the relationships between them that we could identify.

\subsubsection{Change management}
\label{sec:element1}

The change management sub-process comprises \textit{planning} the backsourcing process and \textit{internal communication} of the backsourcing decision to ensure in-house engagement. Elements of this sub-process were described in 15 out of 26 cases, besides the interview study S1, as shown in Table \ref{tab:element1}. Some cases also detailed \textit{post-backsourcing} activities related to this sub process, such as assessing the outcomes of the backsourcing (C3, C11-C14, and C23) and further offshore expansions (C15-C18). 

\subsubsection{Vendor relationship management}
\label{sec:element1b}

Vendor relationship management consists of the activities related to outsourcing \textit{contract termination}, to having a \textit{backsourcing agreement} for handling the backsourcing efforts, and to the possible creation of a \textit{post-backsourcing agreement} to support the parent organization after the backsourcing process has been completed. The backsourcing process requires a high level of interaction with the vendor, and is thus impacted by and affects the client-vendor relationship. The elements related to this sub-process were described in cases C3, C13, C14, C18 and C23 (Table \ref{tab:element1b}).

\begin{table*}[ht]
\def\arraystretch{1.1}
\caption{Vendor relationship management} 
\label{tab:element1b}
\begin{tabular}{P{.1\textwidth}P{0.12\textwidth}P{0.05\textwidth}P{0.63\textwidth}}
\toprule
\textbf{Category} & \textbf{Element} & \textbf{Type} & \textbf{Observations} \\
\midrule
Contract termination & Terminate outsourcing agreement & activity & In all cases, termination of the outsourcing relationship was initiated by the client. According to S1, the termination of the outsourcing agreement triggered a backsourcing decision in many companies. Cases C3 and C18 detailed two distinct options for terminating outsourcing agreements: early and delayed contact termination.\\ \addlinespace
 & Terminate contract early (P5) & practice & In C3, the contract was terminated early to avoid the adverse effects of a relationship breakdown. It required the client to take more responsibility for knowledge transfer. \\ \addlinespace 
 & Delay contract termination (P6) & practice & In C18, a dependency of the vendor on a software release led to the postponement of the outsourcing termination until a new in-house organization was ready. Postponing termination helped the companies in C18 and C23 ensure continuity of operations and allowed them to plan the backsourcing process. \\ \addlinespace 
 & Extend outsourcing contract & artifact & Describes the clauses and conditions that governed the outsourcing agreement. In the case reported in C23, the outsourced contract did not include definite termination clauses. In order to execute the backsourcing decision, the company first renegotiated a contract extension including such clauses. \\
\midrule
Backsourcing agreement & Draw up backsourcing agreement with the vendor & activity & Establishes the vendor's responsibilities for knowledge transfer, as described in cases C3, C13, C14. In C3, relationship breakdown and a lack of exit conditions in the outsourcing agreement led to two months spent negotiating a backsourcing agreement. \\ \addlinespace
 & Backsourcing contract & artifact & A new contract that formalizes the backsourcing agreement and describes roles and responsibilities in the various parts of the backsourcing process. In C13, such a contract outlined the handover date and scope of backsourcing. In the case reported in C23, the original outsourcing contract did not allow the client to bring services back, so the company negotiated an extension including such clauses. \\ \addlinespace
 & Contract early (P7) & practice & Draw up a new contract with the vendor as soon as possible to ensure commitment to the backsourcing process (C13, C14). \\ \addlinespace 
 & Use external experts (P8) & practice & In C3, the organization engaged a specialist contract consultant to ensure the renewal of third-party contracts. \\ 
\midrule 
Post-backsourcing & Maintain a business relationship with vendor (P9) & practice & In C14, the company maintained a relationship with the previous vendor as an IT service provider after contract termination. That was possible due to a positive attitude towards the vendor. \\
\bottomrule
\end{tabular}
\end{table*}

\subsubsection{Competence building}
\label{sec:element2}

This sub-process focuses on acquiring the competence necessary to successfully assume responsibility for the previously outsourced activities. It includes transferring outsourced knowledge to the receiving organization, and building any new competences needed. \textit{Knowledge transfer} needs to deal with both \textit{explicit knowledge}, which is structured and embedded into artifacts, such as source code, data repositories, and documents describing a task or service, and \textit{implicit knowledge}, which is typically embodied in the experience of individuals.

\begin{table*}[htbp]
\def\arraystretch{1.1}
\caption{Competence building}
\label{tab:element2}
\begin{tabular}{P{.1\textwidth}P{0.12\textwidth}P{0.05\textwidth}P{0.63\textwidth}}
\toprule
\textbf{Category} & \textbf{Element} & \textbf{Type} & \textbf{Observations} \\
\midrule
Knowledge transfer planning & Plan knowledge transfer & artifact & In C21, a plan was drawn up to guide the efforts required for knowledge transfer. The plan was grounded in the components of the software development process. In C23, the plan detailed strategies for incremental transfer and selective backsourcing. \\ \addlinespace
 & Responsibilities for knowledge transfer & attribute & Describes the client's and vendor's responsibilities for competence building. In C3, this was a collaboration between both parties. In C13 and C14, the client took responsibility for the migration due to the risk of relationship breakdown. \\ \addlinespace 
 & Backsourced knowledge & artifact & Knowledge transferred back from the vendor (C4-C10, C21, C23). \\ \addlinespace
 & Knowledge symmetry & attribute & Describes the knowledge equivalence between client and vendor. In  C4-C10, the lack of knowledge symmetry imposed additional complications for knowledge transfer. \\ \addlinespace
 & Knowledge type & attribute & Type of knowledge. According to C15-C20, suitable knowledge transfer practices depend on whether knowledge is: \newline
- Explicit: knowledge codified into documents, and thus easily stored and accessed. Knowledge transfer is dependent on the quality of knowledge repositories (C19 and C20) and the infrastructure for maintaining them (C3 and C13). \newline - Implicit: the capability of individuals to provide solutions via skills and competencies (C3, C17, C20, C23). In C15-C18, built through social interaction and practical work. Paper S10 reported that jobs were often transferred back along with the project to cover for relevant implicit knowledge. \\ \addlinespace
 & Task inter-dependency & attribute & Cases C4-C10 used task inter-dependency to analyze how different knowledge-building practices could be used. Decoupled tasks required lower levels of coordination and information exchange than highly coupled ones. C21 and S1 reported difficulties in software development activities due to the high degree of coupling. \\ \addlinespace
 & Task expertise & attribute & Used in C4-C10 to select knowledge building practices. Different expertise levels were required to perform knowledge-building tasks, e.g., specific knowledge depends more on a person's experience and skill, while generic knowledge is easier to grasp. \\
\midrule
Knowledge transfer & Knowledge transfer & activity & The core activity of this sub-process comprises transferring the necessary knowledge from the outsourced environment to the new in-house organization. In C21, the process systematically followed the software development process, recording issues for each activity (e.g., requirements, architecture, implementation, and testing). \\ \addlinespace
 & Layered knowledge transfer (P10) & practice & C23 employed a layering approach (a.k.a. ``peeling the onion'') to transfer knowledge, starting with the components that were easier to tackle, gradually moving up the difficulty ladder towards those that were more difficult and/or central to the functioning of operations. \\ \addlinespace
 & In-house knowledge building practices (P11) & practice & A set of practices supporting the re-acquisition of previously outsourced knowledge. The companies in C4-C10 used the type of knowledge, the expertise required, and task interdependencies to select among the following practices: \newline
- C4 and C10 applied formal processes and tools for work coordination (e.g. scheduling, documents, and formal communications) to organize the transfer of knowledge into the house; \newline
- C5 used an expertise coordination approach, i.e., identifying where task-specific knowledge was needed and where it was located, and bringing them together; \newline
- C6 rebuilt knowledge via ongoing accumulation of experience, i.e., learning by doing; \newline
- C6-C9, and C18, made use of coworking and cooperation with peers to share knowledge among personnel. \\ 
\addlinespace
 & Socialization for knowledge-sharing (P12) & practice & Knowledge-sharing via social interactions brought together experienced employees and novices. In C3, the company conducted technical interviews with experienced personnel to avoid loss of technical knowledge. In C18, training programs fostered knowledge building between co-located units. In C20, the company promoted workshops, forums, and arenas to foster knowledge sharing. \\ 
\midrule
Training & Train skills and competencies & activity & According to C19 and C20, training promoted knowledge-building by providing key skills and capabilities. Training was often reported to focus on the acquisition of technical knowledge, but in C17, continuous training also ensured that the common culture was maintained between on- and offshore sites. According to C4-C10, to make up for missing in-house capabilities, training was integrated into operations. \\ \addlinespace
 & Train a backup person (P13) & practice & A backup person for each position minimized the risk of losing existing knowledge and continuity of services due to staff turnover (C23). \\ \addlinespace 
 & Cross-site collaboration (P14) & practice & In cases C15-C18, coordination strategies were employed to bring co-located people together and facilitate personnel training in an offshore location. In C19 and C20, face-to-face interaction helped transfer services to the in-house location. \\
\midrule
Management & Use external experts (P15) & practice & C14 and C23 reported used of external consultants for technical advice during knowledge transfer. \\
\bottomrule
\end{tabular}
\end{table*}

The sub-process is primarily discussed in cases C3--C10, C13--C14, C21 and C23, and in the included papers S10 and S16. Collaboration with the vendor was a critical factor in ensuring successful \textit{knowledge transfer} in cases C3, C21, and C23. Cases C13 and C14 reported challenges when the vendor did not assist with knowledge transfer and building.

\subsubsection{Organizational build-up}
\label{sec:element3}

Building up the organization that will take over the responsibility for the backsourced activities is vital to the success of backsourcing. During backsourcing, this new organization will absorb the knowledge transferred back in-house and ensure the continuity of services once the transfer is complete. Twelve cases of interest (C3, C11--C18, C21, C20, C23), a survey (S10), and an interview study (S15) provided evidence for this sub-process and its related activities. 

A vital activity concerning \textit{setting up an in-house infrastructure} for development and operations is mentioned in four cases, i.e., C3, C13, C14, and C21. Other activities include maintaining a pool of skilled professionals; this is reported to be accomplished by means of \textit{recruitment} (C3, C11, C23, and S15) \textit{personnel retention} (C3, C15, and C18) and \textit{personnel relocation} (C3, C16--C18, C20 and S10). Finally, combined efforts in recruitment and training ensure in-house capabilities for development and operations (C3, C4--C11, C20, and C23).

\begin{table*}[htbp]
\def\arraystretch{1.1}
\caption{Organizational build-up}
\label{tab:element3}
\begin{tabular}{P{.1\textwidth}P{0.12\textwidth}P{0.05\textwidth}P{0.63\textwidth}}
\toprule
\textbf{Category} & \textbf{Element} & \textbf{Type} & \textbf{Observations} \\
\midrule
Organizational planning & Re-organization plan & artifact & In C11, the company made a plan for rebuilding the internal IS department, including actions such as acquiring resources needed for in-house operations and recruiting developers from the vendor. \\ \addlinespace
 & Set-up in-house environment & activity &
Acquisition, configuration and management of resources for the new in-house environment (C3, C13, C14, C21). Different approaches depending on the resource requirements and environment. E.g., in C14 the company acquired the same hardware as the vendor to build and test an in-house environment. In C3 and C14, existing in-house infrastructure helped the knowledge transfer and reduced reorganization costs. \\ \addlinespace 
& In-house environment & artifact & The technical resources including hardware and software used for the backsourced activities (C13 and C14), and to store the backsourced knowledge (C3 and C13). Supports continuity of services (see Section \ref{sec:element4}). \\ \addlinespace
\midrule
Recruitment & Recruit to cover key skills and competencies & activity & C3, C11, C23, and S15 reported using recruitment to secure skills and competencies for the new in-house organization. Long outsourcing relationships (S15) and high employment turnover (C23) depleted in-house operational capabilities. To ensure that key capabilities were transferred to new recruits, this activity was combined with training (see `Train skills and competencies' in Table \ref{tab:element2}) in S15 and C23. Companies adopted different practices for recruitment, shown below as practices.\\ 
 & Recruit from vendor (P16) & practice & In C3, the new organization managed to retain the vendor's employees temporarily to build up its capabilities for in-house development and operations. In C11 and C23, the client negotiated the transition of its co-located employees back in-house. This practice is also mentioned by participants of a survey study (S10). \\ \addlinespace 
 & Recruit from the job market (P17) & practice & Recruitment campaigns were used in C3 and C15 to hire professionals from the job market. In C15, the responsibility for recruitment was assigned to an experienced leader of the offshore organization. \\ \addlinespace 
 & Short-term recruitment (P18) & practice & Recruit short-term to fill up missing capabilities in the new organization (C14), to support the continuity of operations (C3), or to cover for loss of senior management (C3). In C13, external specialists were recruited as freelancers. \\ 
 \addlinespace
 & Use external recruitment experts (P19) & practice & C23 engaged a transitional vendor to support recruiting and training of personnel for specific areas. \\ 
 & Recruitment requirements & attribute & This characterized the prerequisites of the recruitment activity, such as: formal education (C18), work experience (C16 and C18), and key skills and competencies (C3). C3 and C11 also reported a requirement in terms of the number of people needed. \\
 \addlinespace
 & Recruitment timespan & attribute & According to C3, backsourcing is a timely process, and recruitment strategy can be heavily impacted by time pressure. \\
 \addlinespace
 & Recruitment source & artifact & This indicates a pool of professionals to draw upon for recruitment purposes. Depending on availability, companies recruited professionals from different sources: new employees hired from the job market (C16, C18-C20), existing personnel transferred to the new organization under the backsourcing agreement (C13, C19-C21), and the vendor's employees (C3, C11-C13). \\
\midrule
Personnel retention & Retention of key personnel & activity & This refers to the ability to retain key staff, often related to the needs of implicit knowledge (C3, C15, and C18). \\
 & Improve staff conditions (P20) & practice & Offer better conditions for existing staff as a means of retaining them. In C3, C15, C18, the organization offered incentives for staff to transfer back. \\ 
 \addlinespace
 & Watch employees (P21) & practice & Monitor and control staff involved in backsourcing. To address risks of turnover, C15 and C18 employed control mechanisms, C14 closely monitored all employees involved in the backsourcing process, and C3 established a forum to represent the views of those to be transferred back. \\ 
\midrule
Personnel relocation & Bring outsourced people home (P22) & practice & Bring outsourced staff back in-house to cover for missing capabilities in the new organization and prevent loss of existing knowledge (C3, C20, and S10). \\ 
 \addlinespace
 & Reassign experienced staff (P23) & practice & Transfer existing staff to the new organization. The companies in C16 and C18 transferred experienced developers to facilitate the acquisition of skills in their new offshore subsidiary; C16 also transferred experienced managers to provide leadership. \\
 \addlinespace
 & Stepwise transition of personnel (P24) & practice & In C17, the company had a large pool of developers available, so they employed a stepwise approach, moving one or two of them at a time to the project with proper support and training. This strategy allowed C17 to control the process on an individual level, building technical and domain knowledge. \\ 
\hline
\end{tabular}
\end{table*}

\subsubsection{Transfer of ownership}
\label{sec:element4}

Case C11 reported how the outsourced product or service returned to the new organization after components were transferred in-house. The company now became responsible for managing the newly re-integrated knowledge and ensuring \textit{continuity of services} in-house, integrated with other services already in place (C4-C10). Seven cases of interest (C3, C13, C15-C18, and C23) and two other empirical studies (S1 and S15) provided support for this sub-process.

\begin{table*}[htbp]
\def\arraystretch{1.1}
\caption{Transfer of Ownership}
\label{tab:element4}
\begin{tabular}{P{.1\textwidth}P{0.12\textwidth}P{0.05\textwidth}P{0.63\textwidth}}
\toprule
\textbf{Category} & \textbf{Element} & \textbf{Type} & \textbf{Observations} \\
\midrule
Responsibilities & Take or transfer responsibility & activity & This consisted of assigning responsibility for the backsourced services (entirely or partially) to an organization. In the case of C15, the new organization itself took responsibility for the service. In another case (C23), the company divided responsibility among various offshore organizations. \\ \addlinespace
 & Incremental responsibilities (P25) & practice & The companies in C16 and C18 scaled responsibilities and resources with more commitment from the new organizations. They started on a small scale by allowing one team to collaborate on one product and establish lifecycle management before scaling up. In C16, this strategy was employed with an offshore subsidiary. \\ 
\midrule
Internal business strategy & Role of the new organization & attribute & This describes the role of the new organization at the end of the backsourcing process. In C16, the new organization became an integrated part of the company soon after taking responsibility for the transferred services. In C23, the new organization took the role of an independent service provider. In C3, extra time was required to integrate the newly-built organization with the remainder of the company. \\ \addlinespace
 & Focus on operational continuity (P26) & practice & Reach an agreement with other organizations for continuity of services. In C3, operations were reestablished, but new development activities were suspended while building internal capabilities. \\ \addlinespace 
 & Internal vendor-client model (P27) & practice & In C23, the new organization adopted a vendor-client service model with other organizations in the same company. The model defined goals and metrics through which it monitored the performance of the services. The organization periodically conducted user satisfaction surveys. At the time of the study, the new organization was also planning to establish SLAs to establish formal coordination with other business units.\\ \addlinespace 
\midrule
Continuity of services & Development and operations in-house & activity & This ensures continuity of in-house services once knowledge has been transferred into the new organization (C13 and C14). In C3, operations continued immediately after knowledge transfer, but new software development was postponed due to lack of internal capabilities. In C14, a great deal of effort and a highly motivated staff were required to provide in-house service. Efforts were reduced in C13 due to an existing in-house team already familiar with the backsourced service. \\ \addlinespace
 & Set up coordination and collaboration over sites (P28) & practice & In cases C15-C18, coordination strategies  supported service re-integration when the new organization was an offshore subsidiary. Such practices include:\newline
 - C15-C17 used standardize ways of working and alignment of the software development\newline
 - C16-C17 created virtual teams with developers from different sites\newline
 - C16 and C19 employed face-to-face interaction and peer-to-peer collaboration\newline
 - C17-C18 used exchange visits to build personal relationships and solve problems on site \\ \addlinespace 
\bottomrule
\end{tabular}
\end{table*}

\subsection{Backsourcing Outcomes}
\label{sec:results-outcomes}

Eight out of 17 papers reported the outcomes of the backsourcing process. We classified them into positive and negative, depending on their contribution to the process. Tables \ref{tab:positive-outcomes} and \ref{tab:negative-outcomes} list the outcomes we identified alongside the reported causes or conditions, and the contextual factors that reportedly contributed to achieving the outcomes. Along with the contextual factors, we report related practices (see Section \ref{sec:results-elements}) that were claimed to have contributed to achieving the positive outcomes or mitigating negative outcomes. In the last column, we detail the cases in which we found evidence of the outcome. 

\subsubsection{Positive Outcomes}
\label{sec:positive-outcomes}

According to S14, there is limited evidence in the literature on how to successfully conduct backsourcing. The study aimed to identify success criteria through investigating two backsourcing cases (C13 and C14). The reported factors were: securing the backsourcing process contractually, setting up in-house development environments for close supervision, and hiring knowledgeable and motivated staff. A manager in C14 commented: \textit{``Firstly, a technical in-house team has to be skilled and ready to take on the challenge; secondly, you should secure such transitions with contracts as soon as possible to ensure commitment; and thirdly, systems have to be set in good and clean conditions before migration. Taking over chaos is never a pleasure!''}. C3 reported success criteria collected through interviewee responses: maximizing the number of staff transferred back, ensuring no impact to service continuity, and meeting the transfer handout date. Furthermore, C3 implies that the knowledge transfer activity is crucial for backsourcing success as it `sets the tone' for continuity of services in-house. 

\begin{table*}[htbp]
\def\arraystretch{1.25}
\footnotesize
\caption{Positive outcomes of backsourcing}
\label{tab:positive-outcomes}
\begin{tabular}{P{.17\textwidth}P{.17\textwidth}P{.43\textwidth}P{.14\textwidth}}
\textbf{Positive Outcome} & \textbf{Claimed causes or conditions} & \textbf{Contributing factors} & \textbf{Support} \\
\toprule
Continuity of services & Not rushing to terminate the outsourcing contract & The client and vendor renegotiated a contract extension that allowed the client to plan the backsourcing and build up the organization. A related practice is: P6. & Observation in C23 \\
\midrule
Successful transfer of ownership and responsibilities & New offshore developers integrated with the local team & The company established a new offshore unit and transferred all the products to the new location. Related practices are: P11, P12, and P14 & Observation in C15, C16, and C17 \\
\midrule
Successful re-integration of existing knowledge & Knowledge-building efforts & The companies identified a lack of internal capabilities for operations, and dependencies for specific knowledge. A related practice is: P11. & Observation in C19 and C20, theoretical model in S15 \\
\midrule
Existing knowledge kept & Retain employees and trained backup persons & The company hired employees from the vendor with the skills and capabilities needed to continue operations in-house. Training a backup person with the required skills helped alleviate the challenges of high turnover. Related practices are P14 and P16. & Observation in C23 \\
\midrule
Return of investment in training & Longer retention of employees & The company established a new offshore subsidiary and transferred projects to the new location. Training strategies were in place to further increase the competencies of developers. A related practice is: P14. & Observation in C17 \\
\midrule
Acquisition of skilled professionals & Hire of employees with relevant skills from the vendor & During backsourcing, the client was granted access to the vendor's pool of skilled professionals. Related practice: P16. & Observation in C14 and C23 \\
\addlinespace
& Bringing back outsourced staff & Under the outsourcing agreement, the company in C3 had transferred employees to the vendor, resulting in very low in-house technical capabilities. During the backsourcing, it was necessary to bring those experienced people back. A related practice is: P22. & Observation in C3 \\
& Addressing turnover issues & Companies that build-up the new organization as an offshore subsidiary could experience turnover challenges. To ensure the retention of personnel during the backsourcing process, the companies employed practices such as P14, P20, and P21. & Observation in C15, C17, and C18 \\ 
\midrule
Better service quality & Better coordination with the new subsidiary & The company in C17 acquired an offshore competitor and merged the two organizations. Related practices are: P12, P14, and P24. & Observation in C17 \\
\addlinespace
& Successful recruitment of highly skilled developers & The companies established new offshore subsidiaries, allowing them to recruit from a pool of well-educated and highly-skilled professionals. Improvements in quality were perceived by managers. & Observation in C16 and C17 \\
\midrule
Better control over project priorities & Regained control over operations & Small and medium enterprises (SMEs) contracting large outsourcing vendors had difficulties in receiving priority, due to the size of their contracts. & Observation in C15, C16, and C18 \\ 
\midrule
Lower maintenance costs & Setting up an in-house environment & Under the outsourcing agreement, the vendor could charge for every maintenance task. & Observation in C13 and C22 \\ 
\midrule
Lower recruiting costs & Recruiting and training new personnel via a transitional vendor & A transitional vendor was hired to support the company during the backsourcing process. Through them, the company was able to recruit from a pool of skilled personnel. A related practice is: P19. & Observation in C23 \\
\midrule
Lower operational costs & Internal provision of IS service & Under the outsourcing agreement, the client IS department was transferred to the vendor. Later, the vendor charged increased service fees. This resulted in reduced trust and relationship deterioration. & Observation in C11 and C12 \\ 
\midrule
Market advantages & Cultural proximity between the new organization and target customers & The company established a new offshore subsidiary and transferred projects to the new location. & Observation in C17 \\
\midrule
New business competencies & Hosting and operating the application in-house & The in-house team acquired new technical skills to continue operating. These competencies helped differentiate the company from its competitors. & Observation in C14 \\
\midrule
Expanded offshore business & More responsibilities transferred to an offshore subsidiary & The company established a new offshore organization and transferred projects and employees to the new location. Further expansions added new resources and responsibilities. A related practice is: P25. & Observation in C16 and C18. \\
\bottomrule
\end{tabular}
\end{table*}

In the above-mentioned cases, the perceptions of managers or other stakeholders were used as criteria to assess whether the backsourcing process was successful, and to assess the importance of activities and practices for achieving success. Thus, the criteria we listed above are subjective, based on the experience and perceptions of stakeholders. Other outcomes used to assess benefits of backsourcing include: cost savings by having development or operations in-house rather than the previous outsourcing agreement (C11-C12); or successful completion of a particular activity or sub-process, such as knowledge transfer and building (C19, C20) or continuity of services (C3 and C13-C14).

\subsubsection{Negative Outcomes}
\label{sec:negative-outcomes}

Negative outcomes were described as adverse effects that resulted if risks were not mitigated. In the papers, they were discussed alongside the activities and practices employed to mitigate them (e.g., C3, C13, and C14). Other negative outcomes represent drawbacks such as extra costs (e.g., C17 and C18), time (C3), or effort (C15, C16, and C22) spent on the process. Only one case (C21) reported a failed backsourcing process, caused by failure to re-integrate knowledge in due time. 

\begin{table*}[htbp]
\def\arraystretch{1.25}
\caption{Negative outcomes of backsourcing}
\label{tab:negative-outcomes}
\begin{tabular}{P{.17\textwidth}P{.17\textwidth}P{.43\textwidth}P{.14\textwidth}}
\textbf{Negative Outcome} & \textbf{Claimed causes or conditions} & \textbf{Contributing factors} & \textbf{Support} \\
\toprule
Long time to negotiate a backsourcing agreement & Lack of an exit strategy & The original outsourcing contract lacked clear termination conditions. & Observation in C3 \\
\midrule
Unsuccessful knowledge transfer
& Lack of trust and relationship breakdown & The client-vendor relationship is degraded due to, e.g., low quality of service. A related practice is P6. & Observation in C3, C13, and C14\\ 
\addlinespace
& Dependency on the vendor & The outsourcing contract does not oblige the vendor to be involved in knowledge transfer. & || \\
\addlinespace
& Lack of professionalism & Uncertainty of vendor's staff about keeping their jobs afterwards. & || \\
\midrule
Extra refactoring costs & Restructuring of existing knowledge & Poor service under the backsourcing agreement delivered low-quality software code. & Observation in C15 and C22 \\ 
\midrule
Extra reorganization costs & Merging offshore developers into the new organization & The company established a new offshore subsidiary and transferred projects to the new location. & Observation in C16 \\
\midrule
Extra management costs & Management overhead & The company established a new offshore subsidiary and transferred projects to the new location. & Observation in C17 \\
\midrule
Extra recruitment costs & Hiring and training of new personnel & The company has reduced internal staff since the backsourcing agreement. A related practice is: P17. & Interviewee responses in C3 \\
\midrule
Loss of key staff & Resilient vendor willing to keep staff & The company wants to take advantage of the vendor's technical knowledge and transfer staff back to in-house. A related practice is P16 & Observation in C3 \\
\midrule
Interruption of services & Lack of capabilities in-house & Early termination of the contract did not give the company enough time to build up the new organization. A related practice is P26. & Observation in C3 \\
\midrule
New unit not fully integrated into the organization & Differences in organizational culture & At the time of publication of S8 (one year after knowledge transfer was complete), this was still an ongoing issue. & Observation in C3 \\
\bottomrule
\end{tabular}
\end{table*}
\vspace{-1em}

\subsection{Relationships between backsourcing elements}
\label{sec:results-relationship}

In addition to extracting process elements and outcomes, we aggregated the evidence from the different studies by identifying reported relationships between the items across studies. The type of evidence varied due to different contextual factors and different research approaches. As an example, both cases C3 and C13 provide evidence about the activity \textit{terminate outsourcing contract}. In C3, the context is an \textit{early termination} due to a poor client-vendor relationship, and in C13, a \textit{delayed termination} allowed the company time to build its own internal capabilities. 

When mapping the relationships, we first identified the relationships between elements by analyzing our coding (see Section \ref{sec:results-relationship}). We then selected core elements based on the number of individual relationships. We drew diagrams by arranging elements around a core element and connecting them via the relationships described, thus creating networks. Through successive iterations of the same procedure, we expanded the relationship networks. We choose the networks with the highest number of internal connections and added further detail to them in the form of narrative descriptions, telling a story about the relationship according to the information in the cases. The diagrams represent relationships as the cases report them, and they are not intended to describe a complete view of the backsourcing process or its components. We thus make no other claims with respect to these diagrams than that, to the best of our ability, they try to depict the empirical evidence as we interpreted it in the papers included in this study. We discuss how this work could be expanded by future work in Section \ref{sec:discussion-implications}.

The following subsections describe three relationship networks that we identified by means of this process and found interesting. We illustrate each of them with a diagram and explain how they can be read through narratives that attempt to describe typical process flows. Further, we detail the elements and relationships in the diagram using quotes from the cases. Each diagram includes boxes with different colors representing such elements as reasons for backsourcing, activities, artifacts, attributes, practices, and outcomes. The individual relationships we identified are represented by arrows connecting the elements. The core elements are two activities highlighted in bold in the center of the diagram. A sequence of arrows flowing from top to bottom cross the core elements and highlight a possible reading sequence or flow.

\begin{figure*}[htbp]
 \centering
 \includegraphics[width=1\textwidth]{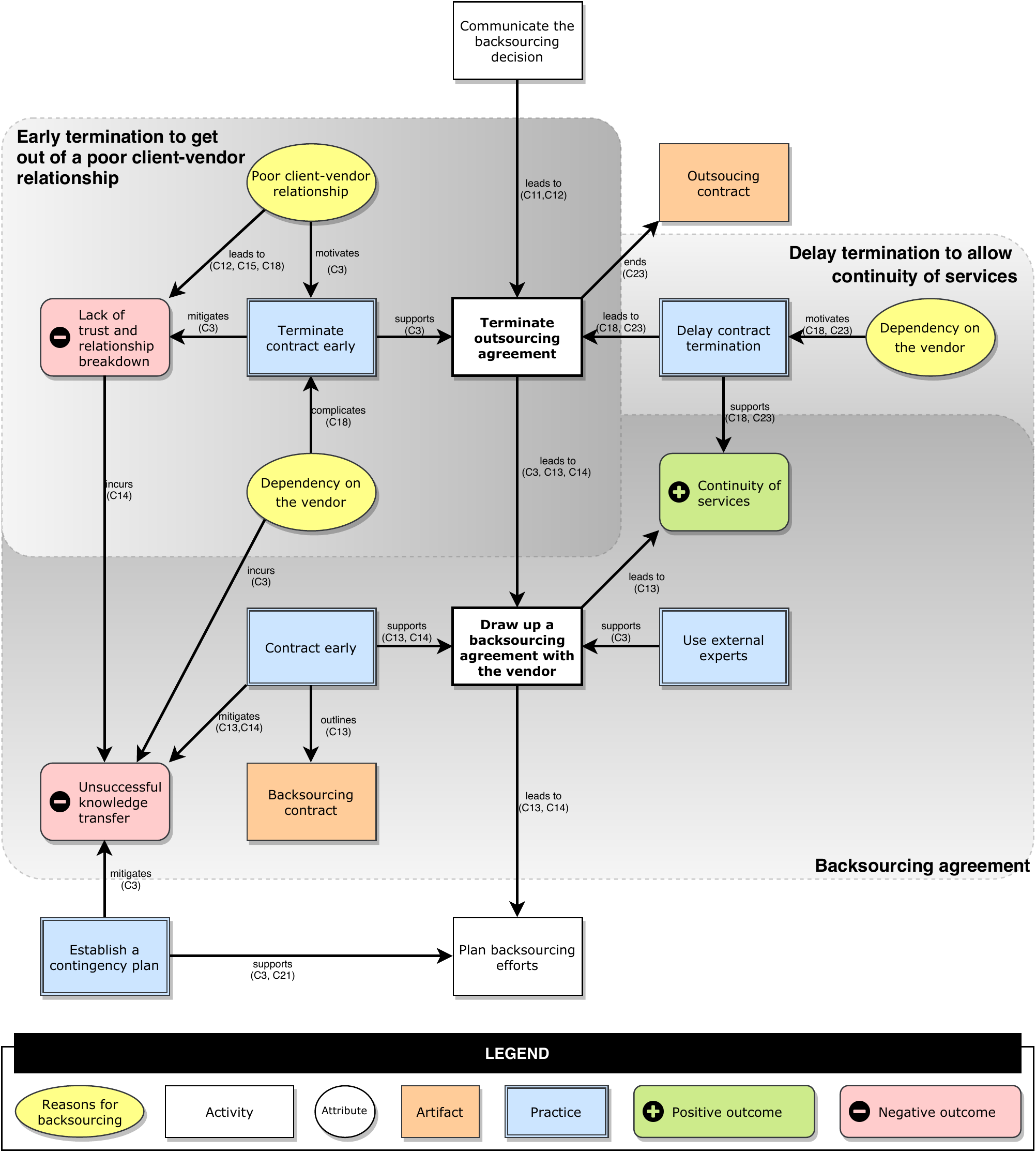}
 \caption{Relationships related to terminating the outsourcing contract and creating a backsourcing agreement.}
 \label{fig:relationship1}
\end{figure*}

\subsubsection{Contractual agreements in backsourcing}
\label{sec:relationship1}

The first relationship is centered on two agreements; the \textit{outsourcing contract} to be terminated; and a \textit{backsourcing agreement} that outlines the responsibilities of the client and vendor during the backsourcing process. We identified evidence of this relationship in cases C3-C4, C11-C15, C18, and C23. In all of them, the decision for backsourcing came from the client. A diagram illustrating this relationship is presented in Figure \ref{fig:relationship1}. 

The basic flow starts with the client deciding and/or \textit{communicating the backsourcing decision}, leading to the \textit{termination of the outsourcing contract} (C11 and C12). Cases C3, C18, and C23 discuss timing issues related to contract termination. Two strategies were identified: 1) \textit{early termination} to end a \textit{poor client-vendor relationship} as soon as possible, and 2) \textit{delayed termination} to guarantee \textit{continuity of services}. In the diagram we show the two options, with their connected elements and relationships, as large gray boxes with dashed borders. After the contract termination, the parties negotiate new conditions to ensure knowledge transfer and continuity of outsourced services back in-house (C3, C13, and C14). Elements and relationships belonging to the new backsourcing agreement are included in the third gray box.

\textbf{Early termination to end a poor client-vendor relationship.} Due to issues with its vendor, the company in case C3 decided to \textit{terminate the outsourcing contract early}. The client set up an arbitrary exit date and kept to the deadline despite requests from the vendor to extend it. The early termination helped avoid adverse effects from \textit{lack of trust and relationship breakdown} due to the \textit{poor client-vendor relationship}. Other cases we identified (C12, C15, and C18) also revealed symptoms of such relationship breakdown due to relationship issues during the outsourcing agreement. 

\textbf{Delay termination to allow continuity of services.} Early termination was not a viable option in C18 and C23 due to \textit{dependency on the vendor}. Instead, the companies in cases C18 and C23 decided to postpone the outsourcing termination to allow the client time to bring the services back and to ensure their \textit{continuity of services} in-house. A managing director in C23 stated, \textit{``We extended the contract for twelve months, just to give us time to get ourselves organized... We tried to resource the whole organization, it was very complex''}. The contract extension also included exit clauses that ensured the outsourcing relationship was eventually terminated after the extension period.

\textbf{Backsourcing agreement.} Following the termination of the outsourcing agreement, the companies in C13 and C14 ensured their backsourcing plan through a \textit{new contract}, mitigating the risks of \textit{unsuccessful knowledge transfer} due to a potential \textit{relationship breakdown}. Due to the circumstances, the company in C3 did not have any exit strategy and ended up spending a long time \textit{drawing up an agreement with the vendor}. As a mitigation practice, the company \textit{established a contingency plan} to speed up knowledge transfer in the event of a complete relationship breakdown. The company in C3 also \textit{used external experts} to support the renegotiation of third-party contracts from the vendor to the client.

\subsubsection{From outsourced knowledge to in-house development and operations}
\label{sec:relationship2}

\textit{Competence building} is a core sub-process of backsourcing, as it describes the transition of knowledge from the vendor to the client and the build-up of necessary in-house knowledge. In this relationship, knowledge comprises both \textit{explicit knowledge}, such as project documentation, guidelines, and scripts; and \textit{implicit knowledge}, e.g., skills, competencies, and experience. This sub-process has the goal of enabling successful \textit{development and operations in-house} (C14 and C16). The elements in this diagram were reported by cases C3-C11, C13-C21, C23, and also by the interview study S1.

\begin{figure*}[htbp]
 \centering
 \includegraphics[width=1\textwidth]{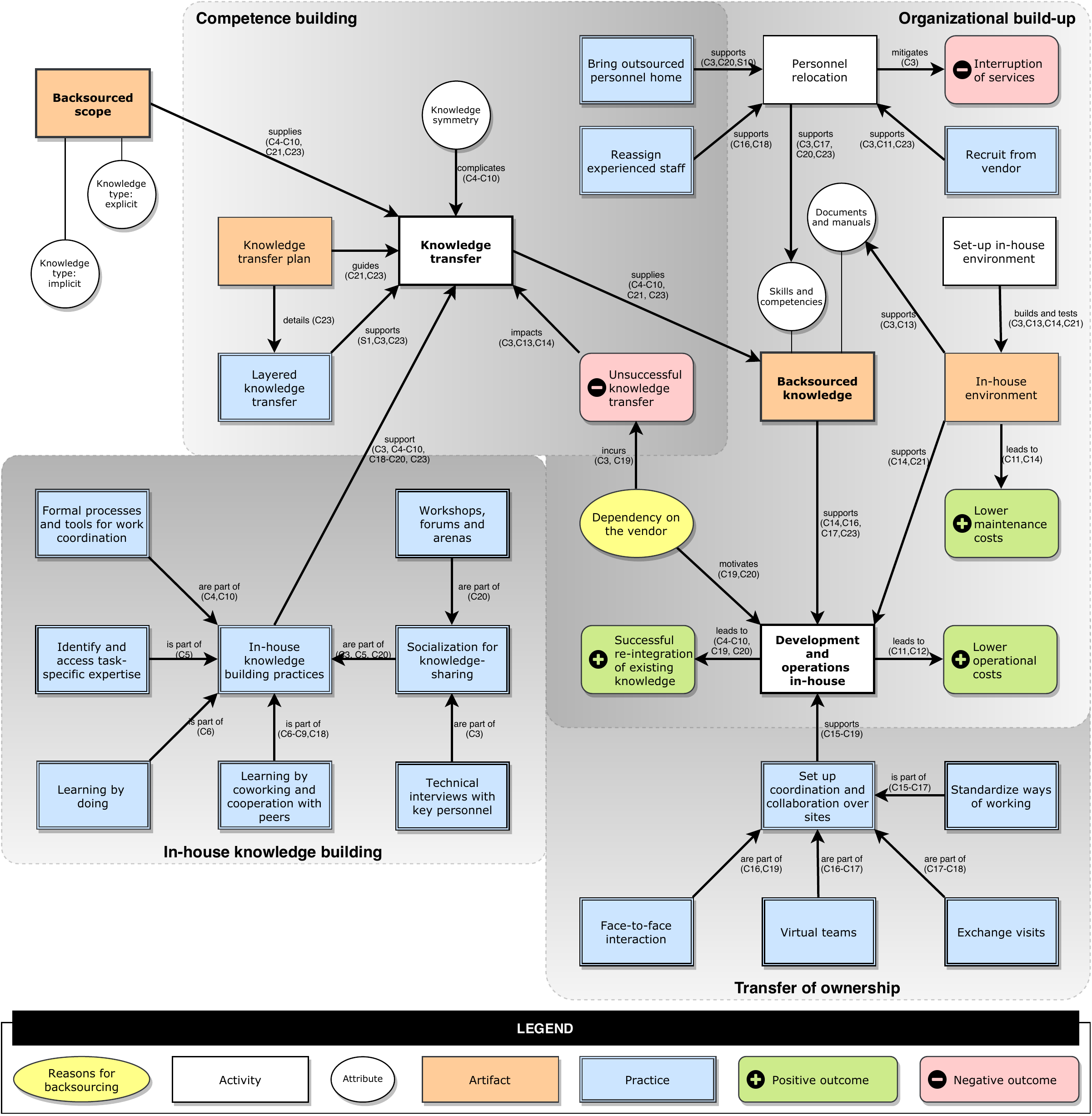}
 \caption{Relationship between knowledge transfer and reestablishing in-house development and operations.}
 \label{fig:relationship2}
\end{figure*}

Figure \ref{fig:relationship2} illustrates four sub-processes: i) \textit{competence building}, ii) \textit{in-house knowledge building}, iii) \textit{organizational build-up}, and iv) \textit{transfer of ownership}. The main flow starts with a relationship between the \textit{backsourced scope} (i.e., outsourced knowledge) and the \textit{knowledge transfer} activity. A set of \textit{in-house knowledge-building practices} supports this activity. A further relationship links \textit{knowledge transfer} and \textit{backsourced knowledge}. Finally, another relationship describes how \textit{backsourced knowledge} provides support to \textit{development and operations in-house}.

\textbf{Competence building.} Cases C15, C19, and C20 describe how knowledge is moved and maintained with the vendor under the outsourcing agreement. This led, in cases C15 and C20, to loss of knowledge on the part of the client. The knowledge difference between the vendor and the client is called \textit{knowledge asymmetry} \cite{ejodame_understanding_2018}. This asymmetry posed a challenge for \textit{knowledge transfer} and motivated researchers to investigate the practices for in-house knowledge building reported in cases C4-C10. 

The companies in cases C21 and C23 employed \textit{knowledge transfer plans} to help lay out the steps and strategies for \textit{knowledge transfer}. In C3, C13, C14, and C19, a \textit{dependency on the vendor} to hand over outsourced knowledge incurred a risk of \textit{unsuccessful knowledge transfer}. A similar dependency motivated C19 and C20 to re-integrate backsourced knowledge and continue operations in-house. As a technology officer in C19 said: \textit{``Instead of becoming dependent on suppliers or, even worse, competitors, we need to protect and share this knowledge within our organization before it gets lost''.}

\textit{Layered knowledge transfer} aimed to organize and prioritize the knowledge to be transferred. In the case of C23, the approach started with the knowledge that was easier to handle and proceeded to the most difficult kind. A managing director in C23 describes this incremental approach as follows: \textit{``We call it peeling the onion. We took the top layer of the onion which was the worst problem and we took care of it and we peel the next layer and the next layer and the next layer and the next until we reach the core''.}

\textbf{In-house knowledge building.} This sub-process comprises a set of practices that provide means of transferring knowledge from the vendor to the client. Note that this is described in Table \ref{sec:element2} as a set of practices associated with the category \textit{knowledge transfer}. We found reports on the use of \textit{in-house knowledge-building practices} in cases C3-C10, C18-C20, and C23. Cases C4-C10 detail a set of characteristics that could affect the adoption of certain knowledge-building practices. The characteristics are the \textit{type of knowledge} (i.e., implicit or explicit), the \textit{task expertise} required, and dependencies, e.g., from other sources of knowledge, or from related tasks. In Figure \ref{fig:relationship2}, we depict the following knowledge-building practices:

\begin{compactitem}
  \item Evidence from two cases (C4 and C10) described how \textit{formal processes and tools} were employed for building \textit{explicit knowledge}. Such methods and tools are the project management or software development methodologies that serve as a template or guide for the service to be backsourced. The context of the two cases was backsourcing of a data processing task; the explicit knowledge, stored as project documentation, guidelines and scripts, described how to conduct such a task.
  
  \item Case C5 described a stepwise practice that consisted of 1) identifying task-related knowledge needs, 2) locating knowledge holders, and 3) bringing them together for knowledge exchange. The situation described in C5 is a financial business process that depended on a series of coupled systems and required coordination between different organizations inside and outside the company. As described by an interviewee in C5: ``There will be some level of interaction; if the person had a credit card, for example, then they'll have to contact [The Company] about it or if they had a particular product they would have to contact the product service about it.''
  
  \item \textit{Learning by doing} and by coworking (described in C6-C9) helped organizations build implicit knowledge through the accumulation of experiences. In C6, the employees designated to operate a fraud audit task learned by redoing the previously outsourced service. In cases C7-C9, coworking and cooperation with peers was used to facilitate learning in the context of backsourcing data center management. The similarities between these four cases lay in the specific expertise required for the task.
  
  \item Social interactions were a means of fostering a culture of knowledge sharing in C3, C5, and C20. They supported building \textit{implicit knowledge} via interaction with peers. This is better explained by a technology officer from C20: \textit{``[…] technical experts tend to keep unique knowledge they do not want to share […] so a main challenge is to change this culture. We had to explain to the guys that the industry is changing […]so they need to understand that they also need to expand their knowledge and combine it with other resources'' (CTO). } Two social interaction practices were reported in the papers: \textit{technical interviews} in C3, and \textit{workshops} in C20. Although we did find supporting evidence for other social interaction formats, such as mentoring programs, we believe they are also applicable in the context of knowledge building. 
\end{compactitem}

\textbf{Organizational build-up.} Evidence gathered from four cases (C3, C13, C14, and C21) describes how the company built up a new organization to handle the backsourced services and to absorb the knowledge to be transferred back. This \textit{backsourced knowledge} supported development and operation in-house. Once this organization was built, in cases C11 and C14, \textit{maintenance costs were lowered}. \textit{Personnel relocation} helped mitigate risks of \textit{interruption of services} due to the loss of key capabilities in C3 and C23. To compensate for a deficit in internal capabilities, companies adopted the following practices: 1) \textit{bring outsourced personnel home}, 2) \textit{reassign experienced staff}, and 3) \textit{recruit people with key skills from the vendor}. A director of operations in C3 stated: \textit{``The biggest hurdle was to make sure that the continuity of service was there. (…) We managed that by hiring a lot of VENDOR's staff who's already in the position.''}

\textbf{Transfer of ownership.} Cases C4, C10, C14, C16, C17, and C23 described the continuity of in-house development and operations once knowledge had been transferred. In the context of C14 and C23, knowledge was defined as the skills and competencies needed for operations, and in C16 and C17, it meant domain knowledge required for software development. In cases C4-C10, the \textit{successful re-integration of existing knowledge} was the main benefit expected from the backsourcing process. C11 and C12 reported another positive outcome: \textit{lower operational costs} by continuing operations in-house once the new organization was built.

In cases where the new organization was an offshore subsidiary (C15-C18), development and operations in-house were supported by \textit{coordination and collaboration over sites}. That comprised a set of practices such as \textit{face-to-face meetings, virtual teams, exchange visits}, and \textit{standardized ways of working}. In the context of cases C15-C18, developers in a new offshore in-house organization employed these practices to share domain knowledge and corporate culture. A manager from the company in C16 explained, \textit{``We had four guys coming here sitting with us for three months... and as they were traveling back to Ukraine they were kind of setting up new teams, being the foundation of the new teams.''} In C19, the face-to-face interaction aimed to locate within the company the existing knowledge needed for the new organization.

\subsubsection{Keeping implicit knowledge in-house}
\label{sec:relationship3}

\begin{figure*}[!ht]
 \centering
 \includegraphics[width=1\textwidth]{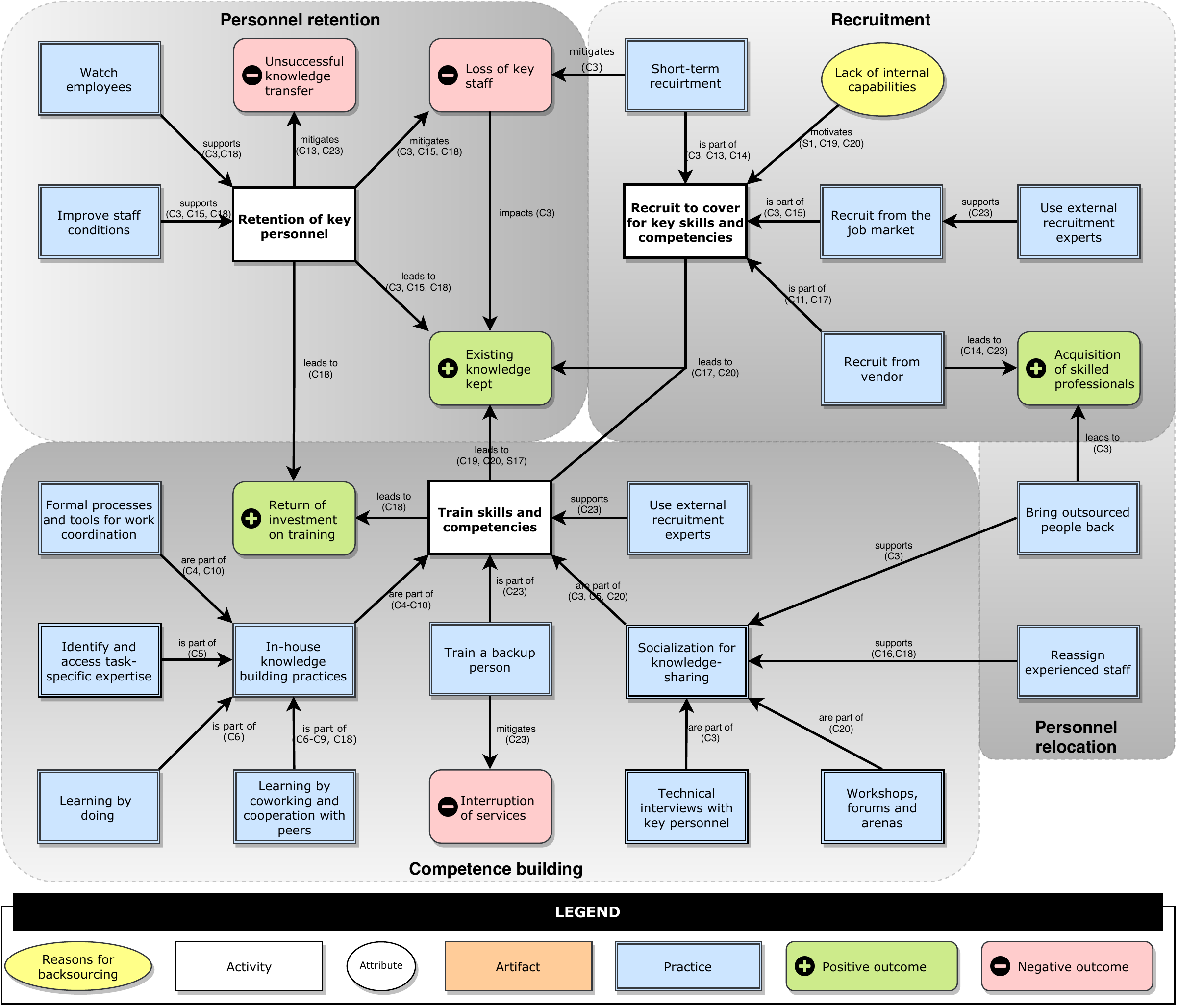}
 \caption{Relationship between sub-processes personnel retention, recruitment, and knowledge building.}
 \label{fig:relationship3}
\end{figure*}

The diagram in Figure \ref{fig:relationship3} describes relationships aimed at ensuring sufficient knowledge to develop and operate software in-house, identified from cases C3-C11, C13-C20, and C23. These relationships are focused on \textit{implicit knowledge} that relies on a person's experience, skills, and competencies. The elements we identified aim to manage personnel to avoid losing existing knowledge, and to build new knowledge based on existing human resources.

One sub-process and two categories are the core elements of this relationship: \textit{competence building} (C19, C20, and S17), \textit{personnel retention} (C3, C15, and C18), and \textit{recruitment} (C3 and C11). The first two contribute directly to the outcome, but recruitment should be aligned with training to contribute to a shared outcome (C17 and C20). Another category, \textit{personnel relocation}, provides practices that support \textit{competence building} (C3 and C18). The structure of the diagram in Figure \ref{fig:relationship3} shows the three core elements draped around the shared outcome \textit{existing knowledge kept}. The reading flow here is not a straight path, but rather an interconnected path that leads to the core outcome.

\textbf{Personnel retention.} A significant risk for backsourcing reported in C3 is the \textit{loss of key staff}, as they possess knowledge that is vital for the new organization. Cases C3, C15, and C18 reported that through \textit{retention of key personnel}, companies mitigated such loss and ensured that \textit{existing knowledge was kept} in-house. C3 reported an attempt by the vendor to cherry-pick people, which meant that the vendor removed people with key competencies from the project. C15 and C18 reported turnover issues among the vendor's staff that could \textit{impact competence building}. Also, risks of \textit{unsuccessful knowledge transfer} were aggravated by uncertainty about job retention among the vendor's staff (C13 and C23). To cope with such issues, companies in cases C3 and C18 closely \textit{watched employees} involved in knowledge transfer; and companies in C3, C15, and C18 offered \textit{improved working conditions} and incentives to address turnover issues. Finally, C18 concluded that retaining employees longer is synergistic with \textit{competence building} (see below), resulting in a \textit{return of investment in training}.

\textbf{Recruitment.} Beside the knowledge obtained by experience, all the skills and competencies needed might not always be found internally (S17 and C23). To make up for this, C3 and C15 reported recruitment campaigns aimed at \textit{hiring professionals from the job market}. The company in C23 had many difficulties in finding the right professionals; thus, they employed \textit{external experts} to help with recruitment and training. As stated by their manager: \textit{``We found the market for those people (i.e., network engineering) very difficult to find. So we went with a vendor that is a transitional vendor. Their modus operandi in life is to help you go from an outsourced to an insourced situation. They hired the people, they trained the people, and at the end of the six-month period, they transferred them to your organization (…).''}

Instead of tapping the job market for the expertise they needed, C11 and C17 \textit{recruited from the vendor's} employees, as they already had expertise and skills in the outsourced service. Another supporting practice was to \textit{recruit short-term} to fill missing capabilities. The company in C3 hired short-term managers to replace key management staff. C13 and C14 reported short-term contracts to cover for lack of technical expertise. As described in C3 and C11, recruitment helped in building capabilities for development and operations.

\textbf{Competence building.} This sub-process helped \textit{keep existing knowledge in-house} (C19, C20, and S17). Most practices related to this activity were presented in section \ref{sec:relationship2} above, so we will not describe them again here. In addition to the practices discussed previously, we identified two that relied on existing personnel: \textit{train a backup person} (C23), and \textit{socialization for knowledge sharing} (C3, C16, C18 and C20). In C23, a backup person was trained for each position; the practice minimized the risk of \textit{interruption of services} due to personnel turnover. Case C3 reported that technical staff from the vendor were brought to the organization for knowledge sharing. Similarly, cases C16 and C18 reported \textit{reassigning experienced staff} to a new offshore organization in an attempt to facilitate \textit{knowledge sharing via social interactions}. C20 employed \textit{workshops} with experienced personnel and novices, resulting in rebuilding of key competencies. The existing knowledge was re-integrated with the new in-house organization, leading to the desired outcome of \textit{keeping existing knowledge in-house}.

%% file: 5-Discussion.tex
\section{Discussion}
\label{sec:discussion}

In this section, we discuss the meaning and relevance of our findings, grouped into four topics. First, we discuss the findings for each research question, and contrast the software development cases with the ones related to other activities. Second, we compare our results with the related secondary studies. Third, we discuss the gaps in and limitations of our results. Finally, we make recommendations for research and practice.

\subsection{Interpretation of the Findings}
\label{sec:discussion-findings}

\textbf{RQ1. What is the context of the reported backsourcing instances?}

We identified 26 backsourcing cases representing a variety of company sizes, business sectors, and backsourced activities. Most organizations built a new organization for handling the backsourced activities onshore, i.e., in the same geographical location as the parent organization.

We did not identify many contextual similarities across cases, except for those reported in the same paper. For example, the only four cases of offshore backsourcing we identified were reported by the same authors \cite{moe_offshore_2012, moe_offshore_2014}. Unfortunately, cases often omitted crucial contextual information, such as details of the outsourced activities or projects (C19-C20, C22, C24-C26), the organization in charge of the service, and the outsourcing contract. The lack of complete contextual information limited our analysis of the backsourcing cases.

Looking at the software development papers, the first one (S14) comparatively explored two cases, one of them related to software development. Papers S5 and S6 reported four cases of failed outsourcing and further backsourcing in Scandinavian companies. The fourth paper (S9) detailed a failed attempt to backsource a software development project of a governmental agency.

Due to the low number of papers and scarce evidence on the backsourcing of software development, we opted to include studies on the backsourcing of other IT-related services such as the hosting of data and applications, and server management. We assumed that such cases would have enough similarities to the software development cases to be worth investigating. The evidence we collected supported that assumption. As an example, the sub-process \textit{organizational build-up} was discussed in twelve cases, half of which were in the context of software development. Many of the activities and practices are supported by evidence aggregated from both contexts.

Besides these six cases, another three (C19, C20, and C22) could have described backsourcing of software development, but they did not explicitly report the backsourced services. Therefore, our analysis considered them among the non-SD cases. Additional evidence about the backsourcing of software development would potentially reinforce current findings and even extend the coverage of the findings to new aspects of backsourcing.

\textbf{RQ2. Why do companies backsource?}

The backsourcing decisions reported were more often reactive than proactive in nature. We found strong links between the reasons \textit{poor-client relationship} and \textit{dependency on the vendor}, with the activity \textit{termination of the outsourcing agreement} supporting this insight. 

The backsourcing decisions were seldom motivated by a single reason. In particular, \textit{poor client-vendor relationship} was often reported alongside other reasons for backsourcing, e.g., \textit{quality problems} and \textit{vendor competence issues}. 

Interestingly, the main reported reasons for backsourcing were strikingly similar to the rationale for outsourcing in the first place \cite{hirschheim_myths_2000}, implying a relationship between failure to meet outsourcing expectations and backsourcing.

Many of the reasons for backsourcing software development were the same as those for backsourcing other IT-related activities. However, none of the software development cases reported \textit{cost savings} as a motivation for the backsourcing; other IT cases extensively reported this reason. It is worth noting that findings from C3 suggest that \textit{cost savings} might not always be reported as an official reason for backsourcing. Thus, it is possible that the software development cases reported only the official reasons and did not investigate more deeply for the true motivation for the backsourcing decision. Furthermore, we did not find any evidence of \textit{changes in strategy or management} in software development cases. This was most probably related to the small number of reported cases, as we could also not identify any explanation for the lack of such changes.

Despite strong arguments in favor of backsourcing, some companies found it difficult to end their poor relationship with the vendor due to \textit{lack of internal capabilities} and \textit{dependency on the vendor}. These reasons against backsourcing often occurred together. In such cases, the decision-making process had to consider the trade-offs between terminating a poor relationship and the need to ensure continuity of the outsourced activities. Companies opting for the former faced the consequences of a relationship breakdown and risks related to knowledge transfer.

Surprisingly, we did not find evidence supporting \textit{dependency on the vendor} in the cases of software development backsourcing. Besides influencing the backsourcing decision, a dependency on the vendor could affect knowledge transfer and in-house development and operations (see e.g., Sections \ref{sec:relationship1} and \ref{sec:relationship2}), but this evidence was gathered solely from cases not related to software development. It is reasonable to assume that the lack of evidence about \textit{cost savings}, \textit{changes in strategy or management}, and \textit{dependencies of the vendor} is due to the limitations and scarcity of the backsourcing cases reported in the literature. Thus, it is worth investigating more practical cases of software development backsourcing with regard to such reasons and their effects on the backsourcing process.

\textbf{RQ3. How do companies backsource?}

While the descriptions of the backsourcing process varied widely in the included papers in respect of both terminology and depth, we were able to identify 13 activities, nine attributes, nine artifacts, and 28 practices related to the process, which we grouped into five sub-processes: \textit{change management}, \textit{vendor relationship management}, \textit{competence building}, \textit{organizational build-up}, and \textit{transfer of ownership}.

Table \ref{tab:process-coverage} shows the contribution provided by each case (plus studies S1 and S10) for the sub-processes and categories we identified. The categories comprise diverse elements (i.e., activities, practices, artifacts and attributes) with a common theme. The first six cases are specific to software development, while the remaining columns list the contributions of other IT-related cases.

\begin{table*}[]
\caption{Contribution of evidence sources to backsourcing process elements}
\label{tab:process-coverage}
\resizebox{\textwidth}{!}{%
\begin{tabular}{@{}lcccccc|cccccccccc@{}}
\toprule
\textbf{Subprocess} & \multicolumn{6}{c}{\textbf{Software development}} & \multicolumn{10}{c}{\textbf{Other IT-related activities}} \\ 
Category  & C13 & C15 & C16 & C17 & C18 & C21 & C3 & C4-10 & C11 & C12 & C14 & C19 & C20 & C23 & S1 & S10 \\ 
\midrule\rowcolor{LightGray} 
\textbf{Change management} & $\bullet$ & $\bullet$ & $\bullet$ & $\bullet$ & $\bullet$ & $\bullet$ & $\bullet$ &  & $\bullet$ & $\bullet$ & $\bullet$ &  & $\bullet$ & $\bullet$ & $\bullet$ &  \\

 Planning & $\bullet$ & $\bullet$ & $\bullet$ & $\bullet$ & $\bullet$ & $\bullet$ &  &  & $\bullet$ &  & $\bullet$ &  & $\bullet$ & $\bullet$ & $\bullet$ &  \\
 Internal communication &  &  &  &  &  &  & $\bullet$ &  & $\bullet$ & $\bullet$ & $\bullet$ &  &  & $\bullet$ & $\bullet$ &  \\
 Post-backsourcing & $\bullet$ & $\bullet$ & $\bullet$ & $\bullet$ & $\bullet$ &  & $\bullet$ &  & $\bullet$ & $\bullet$ & $\bullet$ &  &  & $\bullet$ & $\bullet$ &  \\
\midrule\rowcolor{LightGray} 
\textbf{Vendor relationship management} & $\bullet$ &  &  &  & $\bullet$ &  & $\bullet$ &  &  &  & $\bullet$ &  &  & $\bullet$ & $\bullet$ &  \\
 Contract termination &  &  &  &  & $\bullet$ &  & $\bullet$ &  &  &  &  &  &  & $\bullet$ & $\bullet$ &  \\
 Backsourcing agreement & $\bullet$ &  &  &  &  &  & $\bullet$ &  &  &  & $\bullet$ &  &  & $\bullet$ &  &  \\
 Post-backsourcing &  &  &  &  &  &  &  &  &  &  & $\bullet$ &  &  &  &  &  \\
\midrule\rowcolor{LightGray} 
\textbf{Competence building} & $\bullet$ & $\bullet$ & $\bullet$ & $\bullet$ & $\bullet$ & $\bullet$ & $\bullet$ & $\bullet$ &  &  & $\bullet$ & $\bullet$ & $\bullet$ & $\bullet$ & $\bullet$ & $\bullet$ \\
 Knowledge transfer planning & $\bullet$ & $\bullet$ & $\bullet$ & $\bullet$ & $\bullet$ & $\bullet$ & $\bullet$ & $\bullet$ &  &  & $\bullet$ & $\bullet$ & $\bullet$ & $\bullet$ & $\bullet$ & $\bullet$ \\
 Knowledge transfer &  &  &  &  & $\bullet$ & $\bullet$ & $\bullet$ & $\bullet$ &  &  &  &  & $\bullet$ & $\bullet$ &  &  \\
 Training &  & $\bullet$ & $\bullet$ & $\bullet$ & $\bullet$ &  &  & $\bullet$ &  &  &  & $\bullet$ & $\bullet$ & $\bullet$ &  &  \\
 Management &  &  &  &  &  &  &  &  &  &  & $\bullet$ &  &  & $\bullet$ &  &  \\
\midrule\rowcolor{LightGray} 
\textbf{Organizational build-up} & $\bullet$ & $\bullet$ & $\bullet$ & $\bullet$ & $\bullet$ & $\bullet$ & $\bullet$ &  & $\bullet$ & $\bullet$ & $\bullet$ & $\bullet$ & $\bullet$ & $\bullet$ &  & $\bullet$ \\
 Organizational planning & $\bullet$ &  &  &  &  & $\bullet$ & $\bullet$ &  & $\bullet$ &  & $\bullet$ &  &  &  &  &  \\
 Recruitment & $\bullet$ & $\bullet$ &  & $\bullet$ & $\bullet$ & $\bullet$ & $\bullet$ &  & $\bullet$ & $\bullet$ & $\bullet$ & $\bullet$ & $\bullet$ & $\bullet$ &  & $\bullet$ \\
 Personnel retention &  & $\bullet$ &  &  & $\bullet$ &  & $\bullet$ &  &  &  & $\bullet$ &  &  &  &  &  \\
 Personnel relocation &  &  & $\bullet$ & $\bullet$ & $\bullet$ &  & $\bullet$ &  &  &  &  &  & $\bullet$ &  &  & $\bullet$ \\
\midrule\rowcolor{LightGray} 
\textbf{Transfer of ownership} & $\bullet$ & $\bullet$ & $\bullet$ & $\bullet$ & $\bullet$ &  & $\bullet$ &  &  &  & $\bullet$ &  &  & $\bullet$ &  &  \\
 Responsibilities &  & $\bullet$ & $\bullet$ &  & $\bullet$ &  &  &  &  &  &  &  &  & $\bullet$ &  &  \\
 Internal business strategy &  &  & $\bullet$ &  &  &  & $\bullet$ &  &  &  &  &  &  & $\bullet$ &  &  \\
 Continuity of services & $\bullet$ & $\bullet$ & $\bullet$ & $\bullet$ & $\bullet$ &  & $\bullet$ &  &  &  & $\bullet$ &  &  &  &  &  \\ \bottomrule
\end{tabular}%
}
\end{table*}

Due to the scarcity of the evidence and the quality of reporting, we did not achieve theoretical saturation when coding, and thus find it unlikely that the process elements would add up to a complete and consistent process description. No single case reported all categories, but some (C3, C14, C18, C23) contributed to all five sub-processes, as shown in Table \ref{tab:process-coverage}. The diverse evidence reflects both the differences in reporting, and the fact that the context of the cases varied, implying that backsourcing can be carried out differently in different contexts. Further research is needed to flesh out the process and to better understand the contextual variation.

All sub-processes were supported both by software development and other cases, but not all elements. For example, the practice \textit{use external recruitment experts}, is only described by case C23, not in the context of software development. This practice was aggregated with other recruitment strategies in the sub-process \textit{organizational build-up} that, in turn, is supported by a broader set of backsourcing cases, both SD and other IT-related. Although it is not reported in the context of software development, we see no clear reason why such a practice would not be applicable to software development.

Another example of practices described within a specific study perspective is to \textit{maintain a business relationship with the vendor} after backsourcing, which was reported only in C14. Although reported by an IT case, the context in which this practice occurred is not dependent on the backsourced activity. It depends instead on a positive attitude towards the vendor. Note that C14 did not report dissatisfaction with the vendor as a reason for the backsourcing.

The same logic seems to apply to other elements as well. For example, the activity \textit{internal communication of the backsourcing decision} was supported by S1, C11, C12, and C23, and it is part of the \textit{change management} sub-process that has been extensively reported in both software development and IT contexts. Here too, the lack of evidence does not imply evidence of absence, and we think that this activity may be helpful also in backsourcing software development.

Another interesting pattern concerns two practices: \textit{recruit from the vendor}, and \textit{bring outsourced people home}. Both are unreported in software development contexts. Does that mean that software development companies avoid bringing outsourced people back in-house, either by recruiting or relocation? The evidence does not suggest this (C13, C16, C18 and C21), although there is a lack of detail about how to bring people back in software development cases. Therefore, we draw on the non-SD cases to detail practices P16 and P22.

We found a few activities and practices supported only by software development cases. A noteworthy practice is to \textit{delay the contract termination} to ensure continuity of operations (C18 and C23). Unlike other IT-related services, software development is a process not easily paused and resumed. Iterative processes broadly adopted in outsourced contexts require continuity, and they are dependent on the coordination between distributed teams. This implies additional risks for \textit{competence transfer} and \textit{organizational build-up} that could be mitigated by a late termination, i.e., companies opted to finish building an in-house environment and transfer knowledge before terminating their contracts with the vendor.

We also found that elements associated with a sub-process often interacted with elements from other sub-processes, e.g., by association or dependency, suggesting a degree of overlap instead of a purely sequential process. For example, the activity \textit{development and operations in-house} (Table \ref{tab:element4}) depends on setting up an in-house environment and transferring outsourced knowledge back. Another example is the association between \textit{recruitment} and \textit{training} (see Tables \ref{tab:element2} and \ref{tab:element3}) to cover for key competencies. The findings of such connections revealed interesting relationships between the process elements we investigated further in RQ5.

If the process is not purely sequential, we can assume that there are multiple paths a company can take to successfully carry out backsourcing. Examples of such alternative paths are C3 and C18; the latter ensured complete \textit{knowledge transfer} before \textit{terminating the outsourcing contract}, while the former carried out these two activities in reverse. Furthermore, not all activities and practices are mandatory. For example, a company could complete backsourcing without \textit{recruiting new staff} or \textit{training existing personnel}. However, it is important to note that such activities (and practices) are often related to an outcome, in this case, \textit{acquiring skilled professionals}. 

Ideally, a complete backsourcing process should relate outcomes to each step or action. However, most activities and practices we identified were not explicitly associated with an outcome, which means it is unclear how the actions that were taken during the process contributed to its ends. Even when activities were associated with outcomes, we did not find a common contextual factor across studies to explain the finding. Instead of aiming for a theoretically ideal backsourcing process, we opted to aggregate the available evidence about how backsourcing was conducted. We expanded the elements' description with a narrative of the situations in which they occurred, aiming for a richer depiction of the process.

\textbf{RQ4. What are the reported outcomes of backsourcing?}

Outcomes of the backsourcing process are results, positive or negative, achieved by conducting certain activities and practices rather than by completing backsourcing. 

Most of the outcomes we identified, i.e., 18 out of 23, were reported by only one study. Many were reported by several cases in the same study, resulting from shared data collection and analysis procedures (see e.g. cases C11-C12, C15-C18, and C19-C20). Most observations describing outcomes were reported in C3, C11, C21 and C23. In particular, C3 contributed two thirds of all negative outcomes. The scarce evidence about outcomes suggests that impacts from specific backsourcing actions have not been well investigated, or that they are not easily assessed. 

The majority of observations were drawn from interviews with the participants in the backsourcing cases, and not much is reported about triangulation of these findings with documents or other data sources. In addition, all the cases we identified were post-mortem, i.e., they occurred prior to the investigation. Thus, the evidence supporting the outcomes is grounded in personal opinion and recollection. This calls for further research, aiming to assess the real benefits and drawbacks of backsourcing and its related activities.

Out of the fourteen positive outcomes we identified, seven were reported by software development cases. The most substantial contributors to this were cases C15-C18; all were from the same study. The only positive outcome that reported similar causes for both software development and IT-related cases is \textit{lower maintenance costs}. To achieve this, companies in C13 (SD) and C22 (IT) set up an in-house environment that provided a similar service to their outsourcing vendor without incurring their excessive costs. High costs were reported as a reason for the backsourcing in C22, but not in C13.

Four of the negative outcomes were associated with backsourcing of software development, two of them in aggregation with other IT-related cases: \textit{refactoring costs} and \textit{unsuccessful knowledge transfer}. The former is a consequence of restructuring the low-quality knowledge delivered by the outsourced vendor. The latter was also caused by issues with the vendor, e.g., \textit{lack of trust} and \textit{dependency on the vendor}. The commonalities here are issues with the previous outsourcing that incurred additional challenges during backsourcing. 

It was also noteworthy that the outcomes \textit{continuity of services} and \textit{successfully re-integrated existing knowledge} were not reported in software development cases. Although the evidence across studies is not strong, these outcomes are described as an intended goal in IT-related backsourcing cases. However, they are likely also to be relevant for software development, since \textit{continuity of services} means carrying out the software process in-house after knowledge is backsourced, and \textit{re-integration of knowledge} ensures that domain and technical knowledge is built and maintained by the in-house team. The lack of evidence about such outcomes in the software development context is worth investigating.

\textbf{RQ5. What are the relationships between the context, reasons, processes, and outcomes of backsourcing?}

We identified and described three relationships that present an interesting amalgam of software development and IT backsourcing cases. The connections found are supported by evidence from multiple cases from both contexts. Nonetheless, a few clusters in the diagrams were derived from only one context.

The first relationship (Figure \ref{fig:relationship1}) describes a vital decision point in any backsourcing process: when to \textit{terminate the outsourcing agreement}. There are clear benefits from postponing this activity, and ensuring the vendor's responsibility for knowledge transfer. Nonetheless, companies that had problems with \textit{poor vendor relationships} were willing to terminate the outsourcing agreement as soon as possible. The decision has implications for later steps in the process. Companies opting for \textit{early termination} should acknowledge risks for \textit{unsuccessful knowledge transfer} and \textit{continuity of services}. A candidate for mitigation of this risk was reported by case C13 (SD), in which the company negotiated the vendor's commitment to support the backsourcing process early on.

The second relationship (Figure \ref{fig:relationship2}) describes a long chain of events centered on \textit{knowledge transfer}. This is a crucial activity in backsourcing, and it affects other sub-processes, such as \textit{organizational build-up} and \textit{transfer of ownership}. The issues with knowledge are due to the knowledge type, symmetry, degree of coupling and expertise required. The organization should establish the knowledge needed for \textit{setting up the new in-house environment}. A sloppy execution of this step can lead to problems when \textit{re-integrating knowledge}.

Many connections in the diagram were gathered from an aggregation of software development and IT-related cases. Software development cases contributed mostly with evidence related to \textit{competence building} and \textit{organizational build-up}. In particular, C21 detailed moving a software development project back home using the software process as a guiding structure. Other interesting findings supported exclusively by software development cases are about \textit{setting up coordination and collaboration over sites}; these practices describe how companies in C15-C17 set up a distributed software development environment after backsourcing.

The third relationship (Figure \ref{fig:relationship3}) is centered on the outcome \textit{keeping existing knowledge in-house}. This knowledge is needed for \textit{continuity of services} after backsourcing, and it is found in people's skills and competencies. This implies that people are vital for successful backsourcing due to the \textit{implicit knowledge} they retain. The relationship describes many practices a company can employ to keep implicit knowledge by means of acquiring or retaining key personnel. Some of these practices are complementary, which means they are employed together for greater benefit. A good example is how \textit{relocating experienced personnel from the vendor} and using them to \textit{share knowledge via social interactions} led not only to \textit{keeping existing knowledge}, but spreading it within the new organization (C20).

The \textit{loss of key personnel} (e.g., developers with domain knowledge or technical experts) is a major risk described in this relationship. Cases C15 and C18 reported how to mitigate this risk via \textit{monitoring employees involved in the backsourcing} and \textit{offering better work conditions} as a means of retaining them. These practices were described in the context of software development cases, but we are confident that they are broadly applicable to other contexts. Similarly, cases C16 and C18 provided evidence of strategic \textit{practices for knowledge sharing} that ensure that knowledge is not retained by only a few key people. We venture to say that dependency on key personnel for continuity of services is in many ways as risky as dependence on a vendor.

\subsection{Comparison with Other Reviews of Backsourcing}
\label{sec:discussion-comparison}

We identified five literature studies related to backsourcing that are relevant from our point of view. These papers were excluded as they did not contribute first-hand empirical evidence about backsourcing of software development. However, they are attractive sources for comparing our findings. All five studies included a research question about the reasons for the backsourcing decision (comparable to our research question RQ2), and two of them \cite{veltri_antecedents_2006, wong_jaya_drivers_2008} focused on this question in particular. 

All of the related reviews were systematic literature studies, some with research designs similar to ours. Two papers \cite{von_researchers_2018,wong_jaya_drivers_2008} differ from ours by investigating practitioners' literature (trade journals and press reports) instead of empirical studies. Rather than a simple comparison, their findings could be employed as a complementary interpretation of our findings. The five studies are as follows:

\textbf{Drivers of IT Backsourcing Decision} \cite{wong_jaya_drivers_2008}. The paper presents a literature review of 13 backsourcing cases reported by the press and identifies the main reasons for the backsourcing decisions. Reasons for backsourcing, in order of frequency, are power and politics (6), cost (5), service quality (5), changes in IT role (5), loss of control (4), changes in organizational structure (4), IT resource accessibility (3), changes in strategic directions (1), and changes in vendor organization strategy (1). Interestingly, the most frequent driver, `power and politics at top-level management', does not have a similar reason in our study. Only two cases we identified (C3 and C23) discussed the influence of top management in the backsourcing decision. We suspect this divergence could be due to differences in reporting in the primary sources; ours used empirical studies while their study used press reports. Despite this difference, we noted a few similarities: three reasons for backsourcing (\textit{quality problems}, \textit{cost savings}, and \textit{regain control}) are the topmost cited in both studies. These reasons are part of the `outsourcing expectation gaps' group in Wong and Jaya’s work \cite{wong_jaya_drivers_2008}.

\textbf{Antecedents of information systems backsourcing} \cite{veltri_antecedents_2005, veltri_antecedents_2006}. The second chapter of this report, which is a doctoral thesis, presents a non-systematic literature review on critical factors of the backsourcing decision. The related study differs from ours in several aspects, one being its aim: the literature review provided a theoretical foundation for the thesis, but it is not intended to achieve completeness on the topic. Still, the study reports three main motivations for backsourcing similar to the reasons we identified: \textit{changes in strategy and management}, \textit{cost savings}, and a \textit{poor client-vendor relationship}. The authors analyzed the findings from the perspective of underlying theories, whereas we employed inductive coding based on the narrative from included papers. Thereafter, their work bears no further comparison with ours. The remaining chapters build upon the literature review to develop a theoretical framework focused on the economic, strategic, and relationship considerations of the backsourcing decision. The framework was later applied as a methodological tool to categorize backsourcing cases. 

\textbf{Information system backsourcing, a systematic literature analysis} \cite{leyh_information_2018}. This SLR aimed at answering two research questions similar to ours: what are the drivers of backsourcing (our RQ2), and how do companies backsource their services (our RQ3). The review includes 20 papers; our search strategy also found all of them; however, we excluded 12 of them as they do not present empirical evidence to answer our RQs. Concerning the reasons for backsourcing, the related work identified only three: expectation gaps, internal or external changes, and environmental changes. Expectation gaps are reported most often. Concerning RQ3, the related work described four components of the backsourcing process: (1) transfer and management of knowledge, (2) project management challenges, (3) relationship management, and (4) hiring or re-hiring strategies. Overall, those sub-processes are also confirmed by our SLR, but we extended the process by reporting a further sub-process: \textit{transfer of ownership}.

\textbf{Information systems backsourcing} \cite{von_bary_westner_information_2018}. A literature review that synthesized 31 papers on the topic of backsourcing in IS. The related work differs from ours, mainly by including non-empirical studies (19 out of 31 included papers). Their motivators and decision factors are comparable to our reasons for backsourcing, and once more, `expectation gaps' are ranked highest. The related study distinguishes motivators from decision factors, as the former are triggers and the latter are enablers and barriers to the backsourcing process. Their findings of `implementation success factors' are somehow similar to our outcomes, but we were not able to map them. The related work also derived instructions based on the included papers; we did a similar job by eliciting activities and practices. In our research, we did not find elements (e.g., activities and practices) that are consistently reported, so we instead describe them alongside the context or situation in which they were identified.

\textbf{Gaps between research and practice in the field of information systems backsourcing} \cite{von_researchers_2018}. A review of practitioner literature on information systems (IS) backsourcing. The paper provides an overview of a large set of papers (173) from trade journals and categorizes them according to publication topic and reasons for backsourcing. It also provides a comparison between the practitioner and the academic literature provided by the related work mentioned above \cite{von_bary_westner_information_2018}. The findings suggest that practitioner literature is mostly descriptive, and has limited coverage of topics. As an example, the included papers report only the following reasons for backsourcing: cost savings, quality improvements, and control and flexibility. Once more, the findings support the evidence from academic literature that `expectation gaps' are the reasons most often reported for backsourcing decisions. From a methodological perspective, this related work differs from ours, as we included only empirical evidence, largely limiting practitioner literature. Therefore, we see this work as complementary to ours, and further research ought to verify whether our findings can also be confirmed in cases of practitioners' literature. 

Our work expands on the literature mentioned above by adding a more in-depth description of the elements of the backsourcing process. Only one piece of related work \cite{leyh_information_2018} provided a similar description. However, our work differs from theirs in that we classified the elements according to their contribution to the process (i.e., activities, artifacts, attributes, and practices). We also provided an original contribution by identifying connections between elements that allowed us to describe three relationships. 

\subsection{Research Gaps}
\label{sec:discussion-gaps}

Even though our findings cover a wide range of topics, recurrent themes across the 17 included papers were seldom confirmatory. Two papers could describe different circumstances for the occurrence of the same theme. An example of this issue is the activity \textit{termination of the outsourcing relationship}, which can be either delayed (C23) or hastened (C3). In addition to that, the few cases we identified in the literature are probably a small sample of a large population of backsourcing events. Thus, any interpretation of our findings is an attempt to understand why and how the phenomenon occurs under the reported circumstances, rather than a generalization. We were able to identify several interesting gaps in the literature:

\textbf{Type and strength of evidence.} Most of the evidence we collected is based on observations from a case and from the perspective of an interviewee interpreted by the researcher. However, we managed to gather 71 direct quotes in 7 papers (S2, S4, S5, S7, S13-S15) which provide an unfiltered insight into the data. Besides that, papers often reported how backsourcing cases occurred, and large parts of the articles were dedicated to recounting or narrating the backsourcing story. We noted a lack of in-depth studies that investigate the causes and rationale of the process, or studies that describe the process in detail. Due to this, our study mostly reports patterns we identified and aggregated from the papers, instead of the causal evidence. This gap reveals a need for more experimental and observational studies that could explain the patterns we identified.

\textbf{Lack of vendor’s perspective.} None of the papers identified in our study included the perspective of the vendor company. Some cases (C3, C11-C13) report that the vendor's staff were transferred back in-house, but those people were not interviewed. The client-vendor relationship is often discussed in terms of deterioration and breakdown, especially when detailing the reasons for the backsourcing decision (e.g., C16-C18, and C22). Just one case we identified (C14) reported a business relationship with the vendor after backsourcing. The lack of a vendor's perspective points out a gap in the literature that can be addressed by further studies.

\textbf{Lack of attention to the job market.} Some papers mentioned the potential impact of backsourcing on the local job market, but the information was often conflicting. On the one hand, the survey study in S10 reported no impact on the job market or local economy due to backsourcing; on the other hand, the same paper reports a negative impact on wage rates due to backsourcing. Case C14 reported that due to a global financial crisis, a sizable pool of experienced professionals became available in the job market. The case does not describe the impact on wage rates, but we can assume, based on the supply and demand law, a negative impact due to the influx of new available people. Other papers (S5, S6, and S15) only mentioned that recruitment campaigns drew from the job market, but did not investigate the conditions in which this occurred.

\textbf{Failed backsourcing.} An interesting finding is the only case of failed backsourcing (C21) identified. The paper described an attempt at backsourcing motivated by the need to regain control over an e-government information system. It detailed the actions taken and the technical issues encountered when trying to bring back artifacts of the software development process. The backsourcing process here required much effort, especially in understanding and taking ownership of the outsourced knowledge. The attempt failed due to inability to complete the backsourcing in the required time. It is important to note that in this pilot, the client worked in isolation, without collaboration from the vendor. Despite its negative outcome, C21 provided meaningful evidence regarding the elements and challenges of the backsourcing process. 

\subsection{Implications for Research and Practice}
\label{sec:discussion-implications}

Based on our results, we have a few recommendations for further research and practice. 

One opportunity for further research concerns the relationships illustrated in Section \ref{sec:results-relationship}. The three relationship-diagrams aggregated evidence from multiple papers, but further primary studies are needed to confirm the relationships. It would be valuable to confirm whether our representation appropriately depicts backsourcing events reported from practice. A further study could draw cases from practitioner literature (similar to von Bary et al. \cite{von_researchers_2018}), and use the relationship diagrams to guide qualitative data analysis. Such a study could confirm or provide complementary or conflicting evidence regarding the relationships we identified. In either case, the evidence would strengthen our findings with a deeper understanding of the phenomenon in practice. A limitation of this idea, as already pointed by \cite{von_researchers_2018} is the limited coverage of topics in practitioners' reports.

Other ideas for further research derive from the limitations we discussed in Section \ref{sec:discussion-implications}. First, observational and experimental studies could use the insights we provide in this paper to make testable predictions. Such studies already exist within the topic of outsourcing, e.g., \cite{dutta_offshore_2005, rao_effort_2003, xue_information_2005}, denoting a higher degree of maturity of the field. Furthermore, we encourage researchers willing to conduct case studies of the backsourcing phenomenon to take into account different perspectives such as the vendor, the job market, and failed backsourcing events. Diversity in research views is valuable for further synthesizing of studies. 

From a practitioner’s point of view, our paper could serve as reflexive or instructive reading. Practitioners who have already taken part in a backsourcing process could compare the context, motivations, elements, and outcomes we describe here with their own experiences. We also encourage practitioners to share their views in professional forums, magazines and blogs, thus contributing to further evolution of the field. 

For practitioners who have still to take part in a backsourcing process, or who are currently involved in one, the themes we identify here could be employed as instructional and reference material. Two sections of our results are especially relevant for those planning or currently implementing a backsourcing process. 

First, Section \ref{sec:results-elements} lists key practices that could provide meaningful insights for practical use. Note that the practices were identified in different contextual situations, and one should reflect how they might fit a given real situation. It is also important to consider which practices to adopt based on the expected outcomes. When combined with Section \ref{sec:results-outcomes}, the practices can become a checklist of actions and expected outcomes. We encourage practitioners to evaluate whether the practices led to the desired outcomes. 

Second, the relationship diagrams in Section \ref{sec:results-relationship} may be of interest in providing instructional support when implementing the process. The diagram flow makes explicit the relationships and dependencies between practices and other elements, and the description provides details of the circumstances in which such a relationship occurred in the literature. Most of the circumstances we describe in this paper are based on insights and lessons learned. Additional information about them is found in the included papers; we provided a full list of references in Appendix \ref{sec:appendix1}.

%% file: 6-Conclusions.tex
\section{Conclusions}
\label{sec:conclusions}

This research aimed to understand the phenomenon of backsourcing in information technology. We conducted a systematic literature review, extracting qualitative data from empirical studies reporting mostly real-world experiences. We analyzed the evidence via inductive coding and narrative synthesis. Central findings of our review include the following:

\begin{itemize}
    \item The context of backsourcing cases is assorted. In our study, the backsourcing cases were not limited to a particular business sector, company size, or software activity. Moreover, we did not find patterns in the data related to these contextual factors. Regarding the backsourcing setting, most cases we identified were oriented to the onshore setting, i.e., bringing the service to the same geographical location. However, we noted a few cases of offshore backsourcing.
    
    \item Reasons for backsourcing are often related to a previous outsourcing agreement. The most common reasons for backsourcing were \textit{quality problems}, \textit{high costs}, and \textit{lack of control} of the outsourced services. Often, reasons in favor of backsourcing are aligned to unrealized expectations for the outsourcing agreement. Many companies adopted backsourcing despite considering the reasons against it (e.g., a \textit{dependency on the vendor}) during the decision-making process.
    
    \item There is no commonly-used or streamlined backsourcing process. Each case we identified carried out a specific backsourcing process. The processes comprised several elements, some of which we identified across multiple cases. The evidence revealed five major sub-processes; we also listed activities, attributes, and artifacts that contributed to the sub-processes. We also collected a set of 28 key practices supporting the backsourcing process. We found a few similarities in activities and practices across the cases, but the circumstances in which they occurred were seldom the same.
    
    \item Our results suggest that outcomes are often related to a particular way of conducting a activity or the adoption of a strategic practice. We cannot claim that merely adopting such practices would actually achieve the desired outcomes. However, we believe that adopting practices based on evidence is relevant to the software industry, and that it can foster validation and improvement of such key practices in the field.
\end{itemize}

Based on the evidence we collected about the backsourcing process, we have detailed three relationships of interest. The first relationship describes the roles of contractual agreements in backsourcing, especially the dependencies arising from previous outsourcing contracts. The second describes the flow of knowledge from outsourced setting to the new in-house organization, and the continuity of development and operations. The third establishes a relationship between human resources and keeping knowledge in-house. The relationships are intended to be used as illustrations of a given sub-process flow and to help understand how different parts of the backsourcing process can be implemented.

Finally, we conclude that backsourcing is a complex process; it comprises a plethora of elements, and it depends heavily on circumstantial factors. Therefore, we recommend that companies willing to take this path should first understand the process and its implications. We contribute to this goal by organizing and summarizing evidence from credible cases of backsourcing reported in the literature.

%% file: ms.bbl
\begin{thebibliography}{10}
\providecommand{\url}[1]{#1}
\csname url@samestyle\endcsname
\providecommand{\newblock}{\relax}
\providecommand{\bibinfo}[2]{#2}
\providecommand{\BIBentrySTDinterwordspacing}{\spaceskip=0pt\relax}
\providecommand{\BIBentryALTinterwordstretchfactor}{4}
\providecommand{\BIBentryALTinterwordspacing}{\spaceskip=\fontdimen2\font plus
\BIBentryALTinterwordstretchfactor\fontdimen3\font minus
  \fontdimen4\font\relax}
\providecommand{\BIBforeignlanguage}[2]{{%
\expandafter\ifx\csname l@#1\endcsname\relax
\typeout{** WARNING: IEEEtran.bst: No hyphenation pattern has been}%
\typeout{** loaded for the language `#1'. Using the pattern for}%
\typeout{** the default language instead.}%
\else
\language=\csname l@#1\endcsname
\fi
#2}}
\providecommand{\BIBdecl}{\relax}
\BIBdecl

\bibitem{davis_it_2006}
G.~Davis, P.~Ein-Dor, W.~R~King, and R.~Torkzadeh, ``{IT} offshoring:
  {History}, prospects and challenges,'' \emph{Journal of the Association for
  Information Systems}, vol.~7, no.~1, p.~32, 2006.

\bibitem{dibbern_information_2004}
J.~Dibbern, T.~Goles, R.~A. Hirschheim, and B.~Jayatilaka, ``Information
  {Systems} {Outsourcing}: {A} {Survey} and {Analysis} of the {Literature},''
  \emph{The Data-Base for Advances in Information Systems}, vol.~35, no.~4, pp.
  6--102, 2004, number: 4 Place: New York, NY Publisher: ACM Press.

\bibitem{benaroch_should_2010}
M.~Benaroch, Q.~Dai, and R.~J. Kauffman, ``Should we go our own way?
  {Backsourcing} flexibility in {IT} services contracts,'' \emph{Journal of
  Management Information Systems}, vol.~26, no.~4, pp. 317--358, 2010.

\bibitem{mclaughlin_it_2006}
D.~McLaughlin and J.~Peppard, ``{IT} backsourcing: from ‘make or buy’to
  ‘bringing {IT} back in-house’,'' 2006.

\bibitem{solli-saether_stages--growth_2015}
H.~Solli-Sæther and P.~Gottschalk, ``Stages-of-growth in outsourcing,
  offshoring and backsourcing: back to the future?'' \emph{Journal of Computer
  Information Systems}, vol.~55, no.~2, pp. 88--94, 2015.

\bibitem{herbsleb_global_2001}
J.~D. {Herbsleb} and D.~{Moitra}, ``Global software development,'' \emph{IEEE
  Software}, vol.~18, no.~2, pp. 16--20, 2001.

\bibitem{ebert_global_2016}
\BIBentryALTinterwordspacing
C.~Ebert, M.~Kuhrmann, and R.~Prikladnicki, ``Global software engineering:
  Evolution and trends,'' in \emph{2016 IEEE 11th International Conference on
  Global Software Engineering (ICGSE)}.\hskip 1em plus 0.5em minus 0.4em\relax
  Los Alamitos, CA, USA: IEEE Computer Society, aug 2016, pp. 144--153.
  [Online]. Available:
  \url{https://doi.ieeecomputersociety.org/10.1109/ICGSE.2016.19}
\BIBentrySTDinterwordspacing

\bibitem{sparrow_when_2003}
E.~Sparrow, ``\BIBforeignlanguage{en}{When {Outsourcing} {Fails} to
  {Deliver}},'' in \emph{\BIBforeignlanguage{en}{Successful {IT} {Outsourcing}:
  {From} {Choosing} a {Provider} to {Managing} the {Project}}}, ser.
  Practitioner {Series}, E.~Sparrow, Ed.\hskip 1em plus 0.5em minus 0.4em\relax
  London: Springer, 2003, pp. 195--225.

\bibitem{wong_bringing_2006}
S.~F. Wong, ``Bringing {IT} back home: developing capacity for change,''
  \emph{ICIS 2006 Proceedings}, p.~40, 2006.

\bibitem{cullen_managing_2006}
S.~Cullen, P.~B. Seddon, and L.~Willcocks, \emph{Managing outsourcing: {The}
  lifecycle imperative}.\hskip 1em plus 0.5em minus 0.4em\relax London School
  of Economics and Political Science London, 2006, vol. 139.

\bibitem{whitten_adaptability_2010}
D.~Whitten, ``Adaptability in {IT} {Sourcing}: {The} impact of switching
  costs,'' in \emph{International workshop on global sourcing of information
  technology and business processes}.\hskip 1em plus 0.5em minus 0.4em\relax
  Springer, 2010, pp. 202--216.

\bibitem{brandes_outsourcingsuccess_1997}
H.~Brandes, J.~Lilliecreutz, and S.~Brege,
  ``\BIBforeignlanguage{en}{Outsourcing—success or failure?: {Findings} from
  five case studies},'' \emph{\BIBforeignlanguage{en}{European Journal of
  Purchasing \& Supply Management}}, vol.~3, no.~2, pp. 63--75, Jun. 1997.

\bibitem{gottschalk_critical_2005}
P.~Gottschalk and H.~Solli‐Sæther, ``Critical success factors from {IT}
  outsourcing theories: an empirical study,'' \emph{Industrial Management \&
  Data Systems}, vol. 105, no.~6, pp. 685--702, Jan. 2005, publisher: Emerald
  Group Publishing Limited.

\bibitem{lacity_empirical_1998}
M.~C. Lacity and L.~P. Willcocks, ``An empirical investigation of information
  technology sourcing practices: lessons from experience,'' \emph{MIS
  quarterly}, pp. 363--408, 1998, publisher: JSTOR.

\bibitem{bergkvist_outsourcing_2008}
L.~Bergkvist and O.~Fredriksson, ``Outsourcing {Terms}: {A} {Literature}
  {Review} from an {ISD} {Perspective}.'' in \emph{{ECIS}}, 2008, pp. 458--469.

\bibitem{khan_barriers_2011}
S.~U. Khan, M.~Niazi, and R.~Ahmad, ``Barriers in the selection of offshore
  software development outsourcing vendors: {An} exploratory study using a
  systematic literature review,'' \emph{Information and Software Technology},
  vol.~53, no.~7, pp. 693--706, 2011, publisher: Elsevier.

\bibitem{khan_factors_2011}
------, ``Factors influencing clients in the selection of offshore software
  outsourcing vendors: {An} exploratory study using a systematic literature
  review,'' \emph{Journal of systems and software}, vol.~84, no.~4, pp.
  686--699, 2011, publisher: Elsevier.

\bibitem{smite_empirically_2014}
D.~Šmite, C.~Wohlin, Z.~Galviņa, and R.~Prikladnicki, ``An empirically based
  terminology and taxonomy for global software engineering,'' \emph{Empirical
  Software Engineering}, vol.~19, no.~1, pp. 105--153, 2014.

\bibitem{hirschheim_myths_2000}
R.~Hirschheim and M.~Lacity, ``The myths and realities of information
  technology insourcing,'' \emph{Communications of the ACM}, vol.~43, no.~2,
  pp. 99--107, 2000.

\bibitem{whitten_bringing_2006}
D.~Whitten and D.~Leidner, ``Bringing {IT} back: {An} analysis of the decision
  to backsource or switch vendors,'' \emph{Decision Sciences}, vol.~37, no.~4,
  pp. 605--621, 2006.

\bibitem{von_bary_westner_information_2018}
B.~von Bary and M.~Westner, ``Information systems backsourcing: {A} literature
  review,'' \emph{Journal of Information Technology Management}, vol.~29,
  no.~1, pp. 62--78, 2018.

\bibitem{nujen_backsourcing_2015}
B.~B. Nujen, L.~L. Halse, and H.~Solli-Sæther, ``Backsourcing and {Knowledge}
  {Re}-integration: {A} {Case} {Study},'' in \emph{Advances in {Production}
  {Management} {Systems}: {Innovative} {Production} {Management} {Towards}
  {Sustainable} {Growth}}, ser. {IFIP} {Advances} in {Information} and
  {Communication} {Technology}.\hskip 1em plus 0.5em minus 0.4em\relax Springer
  International Publishing, 2015, pp. 191--198.

\bibitem{nujen_managing_2018}
B.~B. Nujen, L.~L. Halse, R.~Damm, and H.~Gammelsæter, ``Managing reversed
  (global) outsourcing–the role of knowledge, technology and time,''
  \emph{Journal of Manufacturing Technology Management}, vol.~29, no.~4, pp.
  676--698, 2018.

\bibitem{von_bary_how_2018}
B.~von Bary, ``How to bring {IT} home: {Developing} a common terminology to
  compare cases of {IS} {Backsourcing},'' in \emph{Americas Conference on
  Information Systems - AMCIS 2018}.\hskip 1em plus 0.5em minus 0.4em\relax
  USA:press, 2018.

\bibitem{wong_understanding_2008}
S.~F. Wong, ``Understanding {IT} backsourcing decision,'' \emph{PACIS 2008
  Proceedings}, p. 226, 2008.

\bibitem{von_bary_etal_adding_2018}
B.~von Bary, M.~Westner, and S.~Strahringer, ``Adding experts’ perceptions to
  complement existing research on information systems backsourcing,''
  \emph{IJISPM - International Journal of Information Systems and Project
  Management}, vol.~6, no.~4, pp. 17--35, 2018.

\bibitem{lacity_realizing_2007}
M.~Lacity, R.~Hirschheim, and L.~Willcocks,
  ``\BIBforeignlanguage{en}{{Realizing} {Outsourcing} {Expectations}
  {Incredible} {Expectations}, {Credible} {Outcomes}},''
  \emph{\BIBforeignlanguage{en}{Information Systems Management}}, May 2007,
  publisher: Taylor \& Francis Group.

\bibitem{leyh_information_2018}
C.~Leyh, T.~Schäffer, and T.~D. Nguyen, ``Information {System} {Backsourcing}:
  {A} {Systematic} {Literature} {Analysis},'' in \emph{2018 {Federated}
  {Conference} on {Computer} {Science} and {Information} {Systems}
  ({FedCSIS})}.\hskip 1em plus 0.5em minus 0.4em\relax IEEE, 2018, pp. 1--10.
  
\bibitem{wong_jaya_drivers_2008}
S.~F. Wong and P.~Jaya, ``Drivers of {IT} backsourcing decision,''
  \emph{Communications of the IBIMA}, vol.~2, no.~14, pp. 102--108, 2008.

\bibitem{bhagwatwar_considerations_2011}
A.~Bhagwatwar, R.~Hackney, and K.~C. Desouza, ``Considerations for information
  systems “backsourcing”: a framework for knowledge re-integration,''
  \emph{Information Systems Management}, vol.~28, no.~2, pp. 165--173, 2011.

\bibitem{kotlarsky_understanding_2012}
J.~Kotlarsky and L.~Bognar, ``Understanding the process of backsourcing: two
  cases of process and product backsourcing in {Europe},'' \emph{Journal of
  Information Technology Teaching Cases}, vol.~2, no.~2, pp. 79--86, 2012.

\bibitem{veltri_antecedents_2005}
N.~Veltri, ``Antecedents of information systems backsourcing,'' Ph.D.
  dissertation, University of Central Florida, Jan 2005.

\bibitem{veltri_antecedents_2006}
N.~F. Veltri and C.~Saunders, ``Antecedents of information systems
  backsourcing,'' in \emph{Information {Systems} {Outsourcing}}.\hskip 1em plus
  0.5em minus 0.4em\relax Springer, 2006, pp. 83--102.

\bibitem{von_researchers_2018}
B.~von Bary, M.~Westner, S.~Strahringer, ``Do researchers investigate what practitioners deem relevant? Gaps between research and practice in the field of information systems backsourcing,'' in \emph{2018 IEEE 20th conference on business Informatics (CBI)}.\hskip 1em plus 0.5em minus 0.4em\relax IEEE, 2018, pp. 40--49.

\bibitem{kitchenham_evidence-based_2015}
B.~A. Kitchenham, D.~Budgen, and P.~Brereton, \emph{Evidence-based software
  engineering and systematic reviews}.\hskip 1em plus 0.5em minus 0.4em\relax
  CRC press, 2015, vol.~4.

\bibitem{wohlin_guidelines_2014}
C.~Wohlin, ``Guidelines for snowballing in systematic literature studies and a
  replication in software engineering,'' in \emph{Proceedings of the 18th
  International Conference on Evaluation and Assessment in Software
  Engineering}.\hskip 1em plus 0.5em minus 0.4em\relax Citeseer, 2014, p.~38.

\bibitem{cruzes_case_2015}
D.~S. Cruzes, T.~Dybå, P.~Runeson, and M.~Höst,
  ``\BIBforeignlanguage{en}{Case studies synthesis: a thematic, cross-case, and
  narrative synthesis worked example},''
  \emph{\BIBforeignlanguage{en}{Empirical Software Engineering}}, vol.~20,
  no.~6, pp. 1634--1665, Dec. 2015.

\bibitem{miles_qualitative_2014}
M.~B. Miles, A.~M. Huberman, and J.~Saldaña, \emph{Qualitative data analysis:
  {A} methods sourcebook. 3rd}.\hskip 1em plus 0.5em minus 0.4em\relax Thousand
  Oaks, CA: Sage, 2014.

\bibitem{saldana_coding_2015}
J.~Saldaña, \emph{The coding manual for qualitative researchers}.\hskip 1em
  plus 0.5em minus 0.4em\relax Sage, 2015.

\bibitem{bailey_qualitative_2003}
D.~M. Bailey and J.~M. Jackson, ``Qualitative data analysis: {Challenges} and
  dilemmas related to theory and method,'' \emph{American Journal of
  Occupational Therapy}, vol.~57, no.~1, pp. 57--65, 2003, publisher: American
  Occupational Therapy Association.

\bibitem{ampatzoglou_identifying_2019}
A.~Ampatzoglou, S.~Bibi, P.~Avgeriou, M.~Verbeek, and A.~Chatzigeorgiou,
  ``Identifying, categorizing and mitigating threats to validity in software
  engineering secondary studies,'' \emph{Information and Software Technology},
  vol. 106, pp. 201--230, 2019, publisher: Elsevier.

\bibitem{kitchenham_repeatability_2011}
B.~Kitchenham, P.~Brereton, Z.~Li, D.~Budgen, and A.~Burn, ``Repeatability of
  systematic literature reviews,'' in \emph{15th Annual Conference on
  Evaluation \& Assessment in Software Engineering (EASE 2011)}.\hskip 1em plus
  0.5em minus 0.4em\relax IET, 2011, pp. 46--55.
  
\bibitem{aagerfalk2008outsourcing}
P.~J. {\AA}gerfalk and B.~Fitzgerald, ``Outsourcing to an unknown workforce:
  Exploring opensurcing as a global sourcing strategy,'' \emph{MIS quarterly},
  pp. 385--409, 2008.

\bibitem{dutta_offshore_2005}
A.~Dutta and R.~Roy, ``Offshore outsourcing: {A} dynamic causal model of
  counteracting forces,'' \emph{Journal of Management Information Systems},
  vol.~22, no.~2, pp. 15--35, 2005, publisher: Taylor \& Francis.

\bibitem{rao_effort_2003}
B.~S. Rao and N.~L. Sarda, ``Effort drivers in maintenance outsourcing-an
  experiment using taguchi's methodology,'' in \emph{Seventh {European}
  {Conference} {onSoftware} {Maintenance} and {Reengineering}, 2003.
  {Proceedings}.}\hskip 1em plus 0.5em minus 0.4em\relax IEEE, 2003, pp.
  271--280.

\bibitem{xue_information_2005}
Y.~Xue, C.~S. Sankar, and V.~W. Mbarika, ``Information technology outsourcing
  and virtual team,'' \emph{Journal of Computer Information Systems}, vol.~45,
  no.~2, pp. 9--16, 2005, publisher: Taylor \& Francis.

\end{thebibliography}

\begin{thebibliography}{10}
\providecommand{\url}[1]{#1}
\csname url@samestyle\endcsname
\providecommand{\newblock}{\relax}
\providecommand{\bibinfo}[2]{#2}
\providecommand{\BIBentrySTDinterwordspacing}{\spaceskip=0pt\relax}
\providecommand{\BIBentryALTinterwordstretchfactor}{4}
\providecommand{\BIBentryALTinterwordspacing}{\spaceskip=\fontdimen2\font plus
\BIBentryALTinterwordstretchfactor\fontdimen3\font minus
  \fontdimen4\font\relax}
\providecommand{\BIBforeignlanguage}[2]{{%
\expandafter\ifx\csname l@#1\endcsname\relax
\typeout{** WARNING: IEEEtran.bst: No hyphenation pattern has been}%
\typeout{** loaded for the language `#1'. Using the pattern for}%
\typeout{** the default language instead.}%
\else
\language=\csname l@#1\endcsname
\fi
#2}}
\providecommand{\BIBdecl}{\relax}
\BIBdecl

\bibitem{von_bary_etal_adding_2018b}
B.~von Bary, M.~Westner, and S.~Strahringer, ``Adding experts’ perceptions to
  complement existing research on information systems backsourcing,''
  \emph{IJISPM - International Journal of Information Systems and Project
  Management}, vol.~6, no.~4, pp. 17--35, 2018.

\bibitem{wong_bringing_2006b}
S.~F. Wong, ``Bringing {IT} back home: developing capacity for change,''
  \emph{ICIS 2006 Proceedings}, p.~40, 2006.

\bibitem{whitten_bringing_2006b}
D.~Whitten and D.~Leidner, ``Bringing {IT} back: {An} analysis of the decision
  to backsource or switch vendors,'' \emph{Decision Sciences}, vol.~37, no.~4,
  pp. 605--621, 2006.

\bibitem{hirschheim_four_2006}
R.~Hirschheim and M.~C. Lacity, ``Four stories of information systems
  insourcing,'' in \emph{Information systems outsourcing}.\hskip 1em plus 0.5em
  minus 0.4em\relax Springer, 2006, pp. 303--346.

\bibitem{moe_offshore_2014}
N.~B. Moe, D.~Šmite, G.~K. Hanssen, and H.~Barney, ``From offshore outsourcing
  to insourcing and partnerships: four failed outsourcing attempts,''
  \emph{Empirical Software Engineering}, vol.~19, no.~5, pp. 1225--1258, 2014.

\bibitem{moe_offshore_2012}
N.~B. Moe, G.~K. Hanssen, and {others}, ``From offshore outsourcing to offshore
  insourcing: {Three} stories,'' in \emph{2012 {IEEE} {Seventh} {International}
  {Conference} on {Global} {Software} {Engineering}}.\hskip 1em plus 0.5em
  minus 0.4em\relax IEEE, 2012, pp. 1--10.

\bibitem{barney_investigating_2013}
H.~T. Barney, A.~Aurum, G.~C. Low, and K.~Wang, ``Investigating
  post-outsourcing decisions: using the intellectual capital view,'' in
  \emph{Building {Sustainable} {Information} {Systems}}.\hskip 1em plus 0.5em
  minus 0.4em\relax Springer, 2013, pp. 63--75.

\bibitem{butler_isit_2011}
N.~Butler, F.~Slack, and J.~Walton, ``{IS}/{IT} backsourcing-a case of
  outsourcing in reverse?'' in \emph{2011 44th {Hawaii} {International}
  {Conference} on {System} {Sciences}}.\hskip 1em plus 0.5em minus 0.4em\relax
  IEEE, 2011, pp. 1--10.

\bibitem{petalidis_lessons_2018}
N.~Petalidis, ``Lessons from attempting to backsource a government {IT}
  system,'' \emph{Journal of Information Technology Teaching Cases}, vol.~8,
  no.~1, pp. 90--96, 2018.

\bibitem{raghuram_mechanics_2016}
K.~Raghuram, ``Mechanics of making or buying—{Pulling} back the information
  technology engineering in-house,'' in \emph{2016 {International} {Conference}
  on {Electrical}, {Electronics}, and {Optimization} {Techniques}
  ({ICEEOT})}.\hskip 1em plus 0.5em minus 0.4em\relax IEEE, 2016, pp.
  2284--2288.

\bibitem{hirschheim_myths_2000b}
R.~Hirschheim and M.~Lacity, ``The myths and realities of information
  technology insourcing,'' \emph{Communications of the ACM}, vol.~43, no.~2,
  pp. 99--107, 2000.

\bibitem{wong_understanding_2008b}
S.~F. Wong, ``Understanding {IT} backsourcing decision,'' \emph{PACIS 2008
  Proceedings}, p. 226, 2008.

\bibitem{ejodame_understanding_2018}
K.~Ejodame and I.~Oshri, ``Understanding knowledge re-integration in
  backsourcing,'' \emph{Journal of Information Technology}, vol.~33, no.~2, pp.
  136--150, 2018.

\bibitem{kotlarsky_understanding_2012b}
J.~Kotlarsky and L.~Bognar, ``Understanding the process of backsourcing: two
  cases of process and product backsourcing in {Europe},'' \emph{Journal of
  Information Technology Teaching Cases}, vol.~2, no.~2, pp. 79--86, 2012.

\bibitem{nujen_managing_2018b}
B.~B. Nujen, L.~L. Halse, R.~Damm, and H.~Gammelsæter, ``Managing reversed
  (global) outsourcing–the role of knowledge, technology and time,''
  \emph{Journal of Manufacturing Technology Management}, vol.~29, no.~4, pp.
  676--698, 2018.

\bibitem{solli-saether_modenhet_2016}
H.~Solli-Sæther, ``Modenhet i outsourcing, offshoring og backsourcing: tilbake
  til fremtiden?'' \emph{MAGMA}, pp. 48--55, 2016.

\bibitem{aspir_israeli_2019}
T.~Aspir, R.~Gafni, and G.~Gordoni, ``The {Israeli} {CIO}’s journey–{From}
  insourcing to outsourcing and back,'' \emph{Israel Affairs}, pp. 1--19, 2019.

\end{thebibliography}
